\documentclass[aps,pre,preprint,onecolumn,citeautoscript]{revtex4-1}  
\usepackage{amsmath,amssymb,bm} 
\usepackage{graphicx}  
\usepackage[tight]{subfigure}    
\usepackage{color} 
\usepackage[papersize={8.5in,11in}]{geometry}
\usepackage[colorlinks=true]{hyperref}  
\hypersetup{
    bookmarks=true,         
    unicode=false,          
    pdftoolbar=true,        
    pdfmenubar=true,        
    pdffitwindow=false,     
    pdfstartview={FitH},    
    pdftitle={My title},    
    pdfauthor={Author},     
    pdfsubject={Subject},   
    pdfcreator={Creator},   
    pdfproducer={Producer}, 
    pdfkeywords={keyword1} {key2} {key3}, 
    pdfnewwindow=true,      
    colorlinks=true,       
    linkcolor=magenta, 
    citecolor=blue,        
    filecolor=magenta,      
    urlcolor=cyan           
} 

\newcommand{\beq}{\begin{equation}}
\newcommand{\eeq}{\end{equation}}
\newcommand{\nn}{\nonumber \\}
\def\bea{\begin{eqnarray}}
\def\eea{\end{eqnarray}}

\begin{document}
\preprint{NSF-KITP-15-070, arXiv:1506.05111}
\title{Bekenstein-Hawking Entropy and Strange Metals}  
 \author{Subir Sachdev}
 \affiliation{Department of Physics, Harvard University, Cambridge, Massachusetts, 02138, USA}
 \affiliation{Perimeter Institute for Theoretical Physics, Waterloo, Ontario N2L 2Y5, Canada}
\affiliation{Kavli Institute for Theoretical Physics, University of California, Santa Barbara CA 93106-4030, USA}
 \date{June 19, 2015\\
 \vspace{0.6in}}
\begin{abstract} 
We examine models of fermions with infinite-range interactions which realize 
non-Fermi liquids with a continuously variable U(1) charge density $\mathcal{Q}$, and a non-zero entropy density $\mathcal{S}$ at vanishing
temperature.
Real time correlators of operators carrying U(1) charge $q$ at a low temperature $T$ are characterized by a 
$\mathcal{Q}$-dependent frequency  $\omega_{\mathcal{S}} = (q \, T/\hbar)
(\partial \mathcal{S}/\partial{\mathcal{Q}})$ which determines a spectral asymmetry. We show that the correlators match precisely with those
of the AdS$_2$ horizons of extremal charged black holes. On the black hole side, the matching 
employs $\mathcal{S}$ 
as the Bekenstein-Hawking entropy density, and  
the laws of black hole thermodynamics which relate
$(\partial{\mathcal{S}}/\partial{\mathcal{Q}})/(2 \pi)$ to the electric field strength in AdS$_2$.
The fermion model entropy is computed using the microscopic degrees of freedom of a UV complete theory
without supersymmetry.
\end{abstract}
\maketitle 
\section{Introduction}
\label{sec:intro}
 
Holography provides us with powerful tools for investigating models of quantum matter without quasiparticle excitations.
The best understood among these are strongly-coupled conformal field theories (CFTs) in spatial dimensions $d \geq 2$.
Our understanding of such models is built upon the foundation provided by the solvable example of maximally supersymmetric Yang-Mills theory, 
which is known to be holographically dual to string theory on anti-de Sitter space \cite{Maldacena97}. 
 
However, many holographic studies \cite{Hartnoll11,SSarcmp,Iqbal11} 
have focused on experimentally important examples of strongly-coupled quantum matter
which are not CFTs. Of particular interest are compressible states without quasiparticles, or `strange metals', in dimensions $d \geq 2$. 
Broadly defined,
these are quantum states without quasiparticles in which a conserved U(1) charge density $\mathcal{Q}$ can be continuously varied at zero temperature by
a conjugate chemical potential, and the U(1) and translational symmetries are not spontaneously broken.
Solvable examples of strange metals with holographic duals would clearly be of great interest.

Here we consider the strange metal state introduced by Sachdev and Ye \cite{SY92} (SY) in a model of fermions with infinite-range interactions.
The fermion density $\mathcal{Q}$ is conserved and continuously variable, and there is a non-zero entropy density, $\mathcal{S}$, at vanishing 
temperature \cite{PGKS97,GPS00}. The fermion Green's function is momentum independent and so has no Fermi surface (but there is a remnant
of the Luttinger theorem, as discussed in Appendix~\ref{app:nonuniv}). The Green's function is divergent at low frequency 
($\omega$) and temperature ($T$) with a known scaling function
 \cite{PGKS97,PG98,GPS00} (the explicit form is in Eq.~(\ref{GR}) below), 
determined by the fermion scaling dimension, $\Delta$, its U(1) charge $q=1$,
and a spectral asymmetry frequency we shall denote by $\omega_{\mathcal{S}}$. 
This frequency determines the asymmetry between the particle and hole excitations of the non-Fermi liquid. 
The values of $\Delta$ and $q$ are fixed and universal (as in traditional critical phenomena), 
while that of $\omega_{\mathcal{S}}$ varies 
with the compressible density $\mathcal{Q}$ in an apparently non-universal manner. However, the same $\omega_{\mathcal{S}}$,
scaled by the value of $q$, appears
in the correlators of all operators. 

One general way to fix the precise value of $\omega_{\mathcal{S}}$, without a priori knowledge of the
full $\omega$ dependence of the correlator, is the following. The product of the retarded ($G^R$) and advanced ($G^A$) Green's
functions obeys
\beq
G^R (\omega) G^A (\omega)  = \Phi_e \left(\omega - \omega_{\mathcal{S}} \right), \label{ws1}
\eeq
where $\Phi_e (\omega)$ is some {\em even\/} function of $\omega$. So the content of Eq.~(\ref{ws1}) is that $G^R G^A$ becomes an even
function of frequency 
after the frequency shift, $\omega_{\mathcal{S}}$.
With this definition, it was found \cite{PGKS97,PG98,GPS00} that there is a 
surprising general relationship between $\omega_{\mathcal{S}}$ and the 
zero temperature entropy $\mathcal{S}$ density
\beq
\omega_{\mathcal{S}} =  \frac{q\, T}{\hbar} \frac{\partial \mathcal{S}}{\partial \mathcal{Q}}. \label{ws2}
\eeq
Such a relationship was first found in the `multichannel Kondo' problem of a local spin degree of freedom at the
boundary of a CFT2 ({\em i.e.\/} a CFT in 1+1 spacetime dimensions) \cite{PGKS97}. It was later extended \cite{PG98,GPS00} 
to the fermion model of SY, in 
which we define $\mathcal{S}$ and $\mathcal{Q}$ per site, and there are no explicit CFT2 degrees freedom; instead each fermion site
is influenced by a self-consistent environment, and this environment plays a role similar to that of the CFT2 in the Kondo problem.
Both $\mathcal{{S}}$ and $\omega_{\mathcal{S}}$ are non-universal functions of the compressible density $\mathcal{Q}$,
but they are related as in Eq.~(\ref{ws2}).
It is quite remarkable to have a dynamical frequency determined by a thermodynamic property
(which is also defined classically) divided by $\hbar$; other notable instances of connections between observables characterizing low 
frequency dissipation and static thermodynamics or fundamental constants are in Refs.~\onlinecite{damless,kss}.
For the SY state, this value of $\omega_{\mathcal{S}}$ relies on emergent symmetries at low energies, but also requires
careful regularization of the single-site canonical fermions present at high energies. 
In other words, the entropy $\mathcal{S}$ density in Eq.~(\ref{ws2}) counts all the degrees of freedom
in a UV finite fermion model.
Indeed, in this context, Parcollet {\em et al.\/} \cite{PGKS97} note: 
``It is tempting to speculate that a deeper interpretation of these facts is still to be found''.

As we shall demonstrate in this paper, the above properties of the SY state match precisely with 
the quantum theory holographically dual to extremal charged black holes with AdS$_2$ horizons \cite{Romans92,Myers99,Faulkner09,Zaanen}.
As a specific example, we will work with the Einstein-Maxwell theory of (planar or spherical) charged black holes embedded in
asymptotically AdS$_{d+2}$ space, with $d \geq 2$ (the Reissner-Nordstr\"om-AdS solution); however, the key features apply
to a wide class of black hole solutions \cite{Sen05,Sen08,Wald93,Myers93,Wald94,Myers94,Shahin13}.
The correlators of this gravitational theory have the same functional dependence upon $\omega$, $T$, $q$, $\Delta$, and $\omega_{\mathcal{S}}$ as those of the SY state, given in Eq.~(\ref{GR}) below,
and this agreement can be understood by the common conformal and gauge invariances of the two theories \cite{SS10,SS10b,AK15}. 
However, there is a deeper correspondence between the two theories in that Eq.~(\ref{ws2}) for the value of $\omega_{\mathcal{S}}$
also applies in the gravitational theories. 
The holographic computation of correlators yields the value of $\omega_{\mathcal{S}}$
({\em e.g.\/} by using Eq.~(\ref{ws1})), while the right-hand-side of Eq.~(\ref{ws2}) is obtained from a classical gravitational
computation of the Bekenstein-Hawking (or Wald) entropy. The equality in Eq.~(\ref{ws2}) follows from
the classical general relativity of AdS$_2$ horizons of charged black holes (see Section~\ref{blackhole}), and this
potentially provides the 
long sought interpretation for the value of $\omega_{\mathcal{S}}$. 

For a general black hole solution,
the values of $\omega_{\mathcal{S}}$ and $\mathcal{S}$
depend on $\mathcal{Q}$ in a manner different from the SY state, 
but they all continue to obey Eq.~(\ref{ws2}). This difference is not surprising, given that the $\mathcal{Q}$-dependence of $\mathcal{S}$ 
for the SY state 
uses its canonical site-fermion structure in the UV, a characteristic which is not expected to be captured by a gravity dual.
But the validity of Eq.~(\ref{ws2}) in the SY state, and in a wide class of gravity theories, is strong evidence that there is 
a gravity dual which captures all the universal low energy properties of the SY state.

\begin{figure}
\begin{center}
\includegraphics[height=7.5cm]{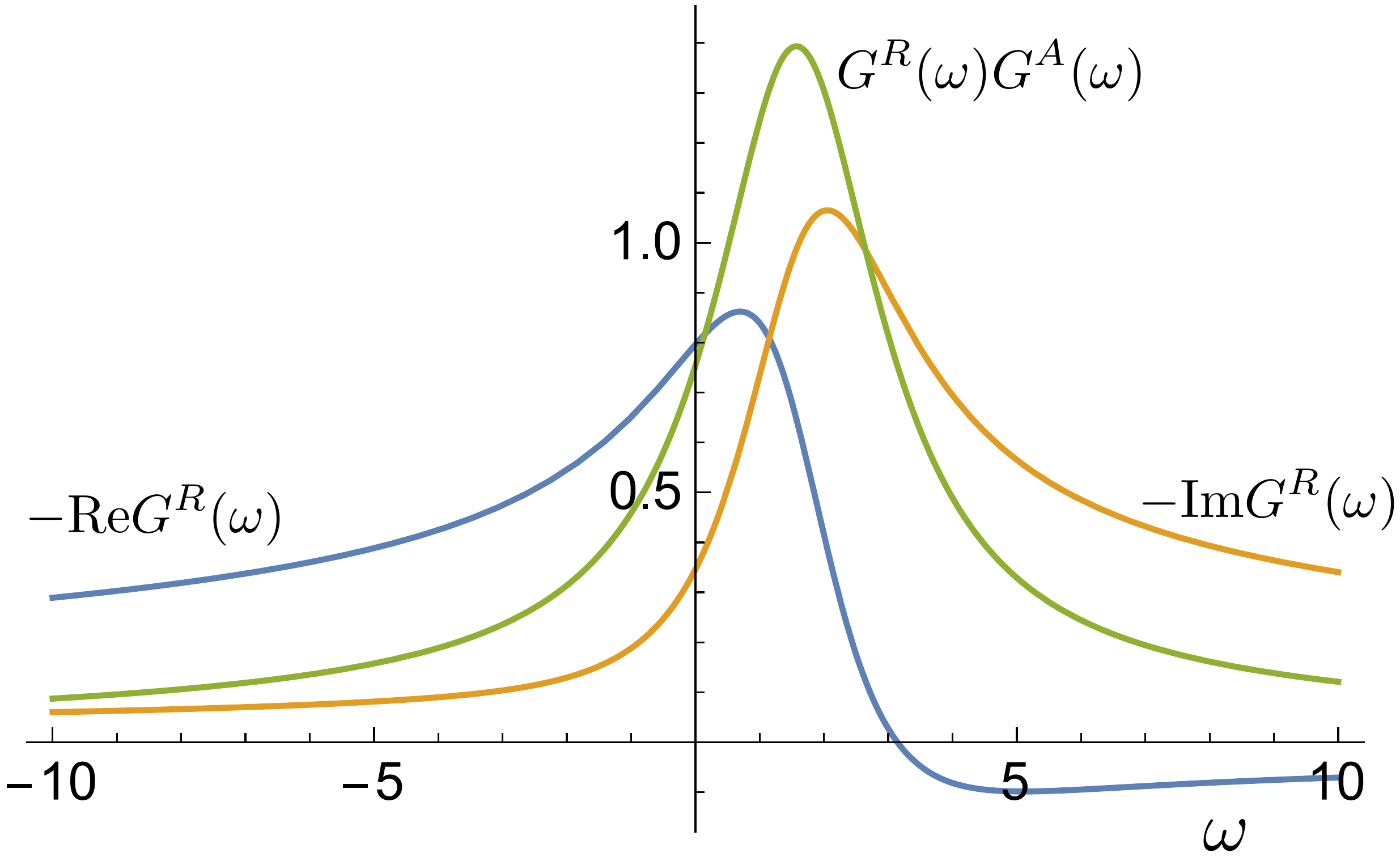}
\end{center}
\caption{Plots of the Green's functions in Eq.~(\ref{GR}) for $
\Delta = 1/4$, $q=1$, $T=1$, $A=1$, $\mathcal{E} = 1/4$ with $\hbar=k_B=1$. 
Note that while neither
$\mbox{Im} G^R (\omega)$ or $\mbox{Re} G^R (\omega)$ have any definite properties under $\omega \leftrightarrow - \omega$, 
the product $G^R (\omega) G^A (\omega)$ becomes an even function of $\omega$ after a shift by $\omega_{\mathcal{S}} = 2 \pi q \mathcal{E} T = \pi/2$.}
\label{Gplot}
\end{figure}

The common two-point correlator of a fermionic operator with U(1) charge $q$, and scaling dimension $\Delta$,
in both the SY and AdS$_2$ theories is \cite{PG98,PGKS97,GPS00,Faulkner09,Iqbal09,Faulkner11}
\beq
G^R (\omega) = G^{A \ast} (\omega) = \frac{-i C e^{-i \theta}}{(2 \pi T)^{1-2 \Delta}}
\frac{\Gamma \left( \displaystyle \Delta - \frac{i  \hbar (\omega - \omega_{\mathcal{S}})}{2 \pi k_B T}  \right)}
{\Gamma \left(  \displaystyle 1 - \Delta - \frac{i \hbar (\omega - \omega_{\mathcal{S}}) }{2 \pi k_B T} \right)}, \label{GR}
\eeq
where $\Delta =1/4$ for the $q=1$ fundamental fermion of the SY state,
the amplitude $C$ is a real and positive, and the angle $-\pi \Delta < \theta < \pi \Delta$ is given by
\beq
e^{2 \pi q \mathcal{E}} = \frac{\sin (\pi \Delta + \theta)}{\sin(\pi \Delta - \theta)}. \label{theta}
\eeq
Here we have found is convenient to introduce a dimensionless, $T$-independent, parameter $\mathcal{E}$ related to $\omega_{\mathcal{S}}$ by
\beq
\mathcal{E} = \frac{1}{2 \pi q} \frac{\hbar \omega_{\mathcal{S}}}{k_B T} \label{SQE1a} 
\eeq
We have therefore introduced three parameters, $\omega_{\mathcal{S}}$, $\mathcal{E}$, and $\theta$,
all of which characterize the spectral asymmetry, and they
can be determined from each other in Eqs.~(\ref{theta}) and (\ref{SQE1a}).
The $T \rightarrow 0$ limit of the Fourier transform of Eq.~(\ref{GR}) shows that $\mathcal{E}$ also defines a `twist'
in the imaginary time fermionic correlator
\beq
G (\tau) \sim \left\{
\begin{array}{ccc} - \, \tau^{-2 \Delta} & , & \tau > 0 \\
e^{-2 \pi q \mathcal{E}} \, |\tau|^{-2 \Delta} &,& \tau < 0 . \label{twist}
\end{array}
\right.
\eeq
It is easy to verify that Eq.~(\ref{GR}) obeys Eq.~(\ref{ws1}).
We show a plot of Eq.~(\ref{GR}) in Fig.~\ref{Gplot} which illustrates the `shift' property of $G^R (\omega) G^A (\omega)$.

\begin{figure}
\begin{center}
\includegraphics[width=7in]{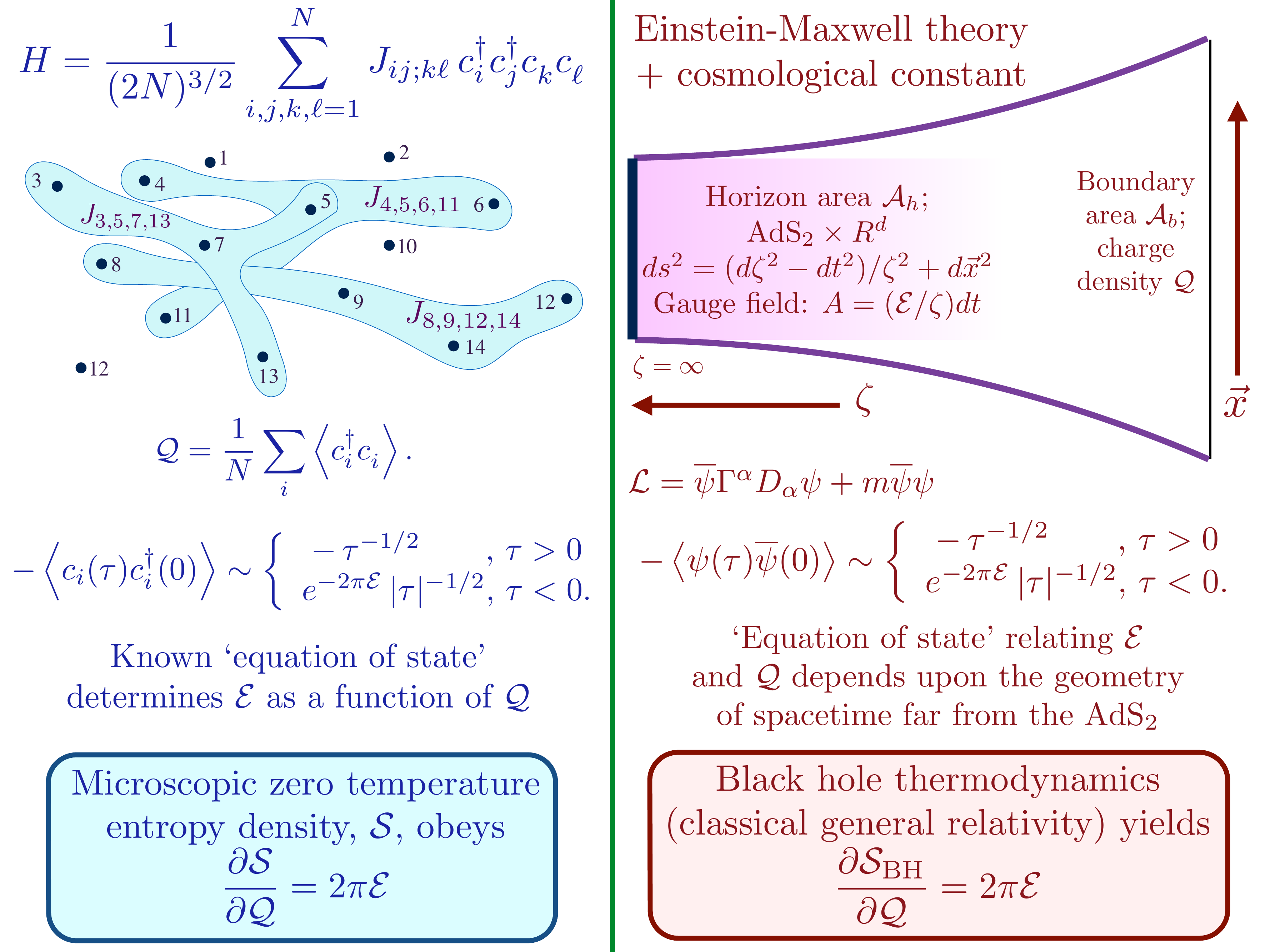}
\end{center}
\caption{Summary of the properties of the SY state (Section~\ref{sec:inf}) and planar 
charged black holes (Section~\ref{sec:holo}) at $T=0$. The spatial co-ordinate $\vec{x}$ has $d$ dimensions. 
All results apply also to spherical black holes considered in Appendix~\ref{app:sphere}.
The AdS$_2 \times R^d$ metric has unimportant prefactors noted in Eq.~(\ref{factor}) which are not displayed above.
The fermion mass $m$ has to be 
adjusted to obtain the displayed power-law. The spectral asymmetry parameter $\mathcal{E}$ appears in the fermion correlators
and in the AdS$_2$ electric field. As the charge $\mathcal{Q}$ is increased, the horizon moves closer to the boundary, and its 
area, $\mathcal{A}_h$, increases. In black hole thermodynamics, the Bekenstein-Hawking entropy density, $\mathcal{S}_{\rm BH}$ 
is related to area of the horizon
via $\mathcal{S}_{\rm BH} = \mathcal{A}_h/(4 G_N \mathcal{A}_b)$, where $G_N$ is Newton's constant.}
\label{fig:summary}
\end{figure}

For the SY state, the previous work \cite{PGKS97,PG98,GPS00} establishes the additional relation in Eq.~(\ref{ws2}), which now relates
the spectral asymmetry parameters $\omega_{\mathcal{S}}$, $\mathcal{E}$, and $\theta$ to $\partial \mathcal{S}/\partial \mathcal{Q}$.
The $T=0$ properties of the SY state reviewed above are summarized in the left panel of Fig.~\ref{fig:summary}.
For our subsequent discussion, it is useful to combine Eqs.~(\ref{ws2}) and (\ref{SQE1a}) in the form
\beq
\frac{\partial \mathcal{S}}{\partial \mathcal{Q}} = 2 \pi \mathcal{E}. \label{SQE1c}
\eeq

In the holographic computation of Eq.~(\ref{GR}), the temperature, $T$, is the Hawking temperature of the black hole horizon \cite{Gibbons77},
and the dimensionless spectral asymmetry parameter $\mathcal{E}$ appearing in Eq.~(\ref{twist}) (and Eq.~(\ref{GR})) 
is determined by the strength of the 
electric field (see Eq.~(\ref{Azeta})) supporting the near-horizon AdS$_2$ geometry \cite{Hartman08,Faulkner09,Faulkner11} (see Fig.~\ref{fig:summary}).
A key observation in the holographic framework is that $\mathcal{E}$, now related to the
electric field, obeys an important identity which follows from the laws of black hole thermodynamics \cite{Bardeen73}
(see Fig.~\ref{fig:summary}; we set $\hbar=k_B = 1$ in the remaining discussion):
\beq
\frac{\partial \mathcal{S}_{\rm BH}}{\partial \mathcal{Q}} = 2 \pi \mathcal{E}, \label{SQE1b}
\eeq
where $\mathcal{S}_{\rm BH}$ is the Bekenstein-Hawking entropy densiy of the AdS$_2$ horizon.
Indeed, Eq.~(\ref{SQE1b}) is a general consequence of the classical Maxwell and Einstein equations, and the conformal
invariance of the AdS$_2$ horizon, as we shall show in Section~\ref{blackhole}.
Moreover, a Legendre transform of the identity in Eq.~(\ref{SQE1b}) was established by Sen
 \cite{Sen05,Sen08} for a wide class of theories
of gravity in the Wald formalism \cite{Wald93,Myers93,Wald94,Myers94,Shahin13}, in which $\mathcal{S}_{\rm BH}$ is generalized to
the Wald entropy.

The main result of this paper is the identical forms of the relationship 
Eq.~(\ref{SQE1c}) for the statistical entropy of the SY state, and Eq.~(\ref{SQE1b}) for the Bekenstein-Hawking
entropy of AdS$_2$ horizons. 
This result is strong evidence that there is a gravity dual of the SY state with a AdS$_2$ horizon.
Conversely, assuming the existence of a gravity dual, Eqs.~(\ref{SQE1c}) and (\ref{SQE1b}) show that such a correspondence
is consistent only if the black hole entropy has the Bekenstein-Hawking value, and endow the black hole entropy 
with a statistical interpretation \cite{ASCV}.

It is important to keep in mind that (as we mentioned earlier) 
the models considered here have a different `equation of state' relating $\mathcal{E}$ to $\mathcal{Q}$: 
this is specified for the SY state in Eq.~(\ref{QE}), for the planar black hole in
Eq.~(\ref{SQE4}), and for the spherical black hole in Eq.~(\ref{s4}).

The holographic link between the SY state and the AdS$_2$ horizons of charged black branes has been conjectured earlier \cite{SS10,SS10b,AK15}, based upon the presence of a non-vanishing zero temperature entropy density and the conformal structure of correlators. 
The results above sharpen this link by establishing a precise
quantitative connection for the Bekenstein-Hawking entropy \cite{JDB73,SWH75} of the black hole with the UV complete computation on 
the microscopic degrees of freedom of the SY state.

It also worthwhile to note here that in the usual matrix large $M$ limit of the AdS/CFT correspondence \cite{Maldacena97},
$\mathcal{S}$ and $\mathcal{Q}$ are both of order $M^2$ \cite{Faulkner09}. So  $\omega_{\mathcal{S}}$ and $\mathcal{E}$
both remain of order unity in this limit. 

We will present an infinite range model and its solution in Section~\ref{sec:inf}. An important result here is the emergent
conformal and gauge invariance in Eq.~(\ref{eq:conf}), which strongly constrains the low energy theory.
We then turn to the Einstein-Maxwell theory of charged horizons in Section~\ref{sec:holo} and show that it also obeys Eqs.~(\ref{ws1}--\ref{SQE1b}).
We conclude with a discussion of broader implications is in Section~\ref{sec:conc}.

\section{Infinite range model}
\label{sec:inf}

SY considered a model of SU($M$) spins with Gaussian random exchange interactions between any pair of 
$N$ sites, followed by the double limit $N \rightarrow \infty$ and then $M \rightarrow \infty$. 
Their Hamiltonian is
\beq
H = \frac{1}{(NM)^{1/2}} \sum_{i,j=1}^N \sum_{\alpha,\beta=1}^M J_{ij} c_{i \alpha}^\dagger c_{i \beta} c_{j \beta}^\dagger c_{j \alpha},
\label{HSY}
\eeq
where the $c_{i \alpha}$ are canonical fermions obeying
\beq
c_{i\alpha} c_{j\beta} + c_{j \beta} c_{i \alpha} = 0 \quad, \quad c_{i \alpha}^{\vphantom \dagger} c_{j \beta}^\dagger + c_{j \beta}^\dagger 
c_{i \alpha}^{\vphantom \dagger} = \delta_{ij} \delta_{\alpha\beta}, \label{antiSY}
\eeq
and there is a fermion number constraint
\beq
\frac{1}{M} \sum_\alpha c_{i \alpha}^\dagger c_{i \alpha} = \mathcal{Q}, \label{defQSY}
\eeq
on every site $i$, with $0 < \mathcal{Q} < 1$. The exchange interactions $J_{ij}$ are independent 
Gaussian random numbers with zero mean and equal variance.

Kitaev \cite{AK15} has recently pointed out that the SY state can also be realized in a simpler model of Majorana fermions 
in which only a single large $N$ limit needs
to be taken, and which also suppresses spin-glass order \cite{GPS99,GPS00,MJR02,MJR03}. 
We will present our results here using a complex fermion generalization of Kitaev's proposal, but we emphasize that essentially
all results below apply equally to the original model of SY in Eq.~(\ref{HSY}).
We consider the Hamiltonian of spinless fermions $c_i$
\beq
H = \frac{1}{(2 N)^{3/2}} \sum_{i,j,k,\ell=1}^N J_{ij;k\ell} \, c_i^\dagger c_j^\dagger c_k^{\vphantom \dagger} c_\ell^{\vphantom \dagger} 
- \mu \sum_{i} c_i^\dagger c_i^{\vphantom \dagger}, \label{H}
\eeq
with
\beq
c_i c_j + c_j c_i = 0 \quad, \quad c_i^{\vphantom \dagger} c_j^\dagger + c_j^\dagger c_i^{\vphantom \dagger} = \delta_{ij}, \label{anti}
\eeq
and the $J_{ij;k\ell}$ are complex, independent Gaussian random couplings with zero mean obeying
\bea
J_{ji;k\ell} = - J_{ij;k\ell} \quad , \quad
J_{ij;\ell k} &=& - J_{ij;k\ell} \quad , \quad
J_{k\ell;ij} = J_{ij;k\ell}^\ast \nn
\overline{|J_{ij;k\ell}|^2} &=& J^2.
\eea
Because there is only a fermion interaction term in $H$, and no fermion hopping, Eq.~(\ref{H}) can be viewed as `matrix model'
on Fock space, with a dimension which is exponentially large $N$.
The conserved U(1) density, $\mathcal{Q}$ is now related to the average fermion number by
\beq
\mathcal{Q} = \frac{1}{N} \sum_i \left\langle c_i^\dagger c_i^{\vphantom \dagger} \right\rangle, \label{defQ}
\eeq
which replaces the on-site constraint in Eq.~(\ref{defQSY}).
The value of $0<\mathcal{Q}<1$ can be varied by the chemical potential $\mu$. The solution described below
applies for any $\mu$, and so realizes a compressible state.

Note that we could equally have defined $\mathcal{Q}$ without the
$1/N$ prefactor in Eq.~(\ref{defQ}); then we would have to define $\mathcal{S}$ as the 
total entropy, and both $\mathcal{Q}$ and $\mathcal{S}$ would be proportional to $N$ in the large $N$ limit. 
The latter scaling would then be similar to the $M^2$ scaling of these quantities in the usual matrix large $M$ limit of the AdS/CFT correspondence \cite{Maldacena97}. But we choose to work here with an intensive definition of $\mathcal{Q}$ and $\mathcal{S}$, and keep the
$1/N$ in Eq.~(\ref{defQ}).

Introducing replicas $c_{ia}$, with $a=1 \ldots n$, we can average over disorder and obtain the replicated imaginary time ($\tau$) action
\beq
S =\sum_{ia}  \int_0^{1/T} d \tau c_{ia}^\dagger \left( \frac{\partial}{\partial \tau} - \mu \right) c_{ia} 
- \frac{J^2}{4N^3} \sum_{ab} \int_0^{1/T} d \tau d \tau'
\left| \sum_i c_{ia}^\dagger (\tau) c_{i b}^{\vphantom\dagger} (\tau') \right|^4 ;
\eeq
(here we are neglecting normal-ordering corrections which vanish as $N \rightarrow \infty$).
Following SY, we decouple the interaction by two successive Hubbard-Stratonovich transformations.
First, we introduce the real field $Q_{ab} (\tau, \tau') $ obeying
\beq
Q_{ab} (\tau, \tau') = Q_{ba} (\tau', \tau).
\eeq
In terms of this field
\bea
S &=& \sum_{ia}  \int_0^{1/T} d \tau c_{ia}^\dagger \left( \frac{\partial}{\partial \tau} - \mu \right) c_{ia} 
+ \sum_{ab} \int_0^{1/T} d \tau d \tau' \Biggl\{ \frac{N}{4J^2} \left[ Q_{ab} (\tau, \tau') \right]^2 \nn
&~&~~~~~~~~~~~~~~~~~ - \frac{1}{2N}
Q_{ab} (\tau, \tau') \left| \sum_i c_{ia}^\dagger (\tau) c_{i b}^{\vphantom\dagger} (\tau') \right|^2 \Biggr\}.
\eea
A second decoupling with the complex field $P_{ab} (\tau, \tau') $ obeying
\beq
P_{ab} (\tau, \tau') = P_{ba}^\ast (\tau', \tau)
\eeq
yields
\bea
S &=& \sum_{ia}  \int_0^{1/T} d \tau c_{ia}^\dagger \left( \frac{\partial}{\partial \tau} - \mu \right) c_{ia} 
+ \sum_{ab} \int_0^{1/T} d \tau d \tau' \Biggl\{ \frac{N}{4J^2} \left[ Q_{ab} (\tau, \tau') \right]^2  +
\frac{N}{2} Q_{ab} (\tau, \tau') \left|P_{ab} (\tau, \tau')\right|^2 
\nn 
&~&~~~~~~~~~~~~~~~~~~~~ -  Q_{ab} (\tau, \tau') P_{ba} (\tau', \tau)
 \sum_i c_{ia}^\dagger (\tau) c_{i b}^{\vphantom\dagger} (\tau')  \Biggr\}
\eea
Now we can integrate out the fermions and obtain an action which can be solved in the saddle-point approximation in the limit of large $N$.
The saddle-point equations are
\bea 
P_{ab} (\tau, \tau') &=&  \left\langle c_{a}^\dagger (\tau) c_{ b}^{\vphantom\dagger} (\tau') \right\rangle \nn
Q_{ab} (\tau, \tau') &=&  J^2 \left| P_{ab} (\tau, \tau') \right|^2
\eea
Notice we have dropped the site index on the fermions, because all sites are equivalent and the saddle-point equations are defined
as a single-site problem.

We do not expect spin-glass solutions in this model, and so we restrict our attention to replica diagonal solutions in which
\beq
P_{ab} (\tau, \tau') = \delta_{ab} G(\tau' - \tau),
\eeq
where $G(\tau)$ is the usual fermion Green's function. In the operator formalism for the underlying Hamiltonian, 
this Green's function is defined in Euclidean time by
\beq
G(\tau_1, \tau_2) = -\frac{1}{N} \sum_i \left\langle T_\tau \left( c_i (\tau_1) c_i^\dagger (\tau_2) \right) \right\rangle
\eeq
where $T_\tau$ denotes time-ordering, and $G(\tau_1-\tau_2) = G(\tau_1, \tau_2)$.
Now the large $N$ saddle-point equations become \cite{SY92}
\beq
G(i \omega_n) = \frac{1}{i \omega_n + \mu - \Sigma (i\omega_n)} \quad, \quad \Sigma (\tau) = -  J^2 G^2 (\tau) G(-\tau), \label{eq:SY}
\eeq
where $\omega_n$ is a Matsubara frequency.  Although they are innocuously simple in appearance, these equations contain a great deal of emergent
scaling structure. 

In the low energy scaling limit, $\omega, T \ll J$, the $i \omega_n + \mu - \Sigma(i \omega_n=0)$ is irrelevant \cite{SY92}. Then,
it is useful to write these equations in imaginary (Euclidean) time, separating the two time arguments of the Green's functions (the self energy has
the value of $\Sigma(i \omega_n = 0)$ subtracted out):
\bea
\int d \tau_2 G ( \tau_1, \tau_2 ) \Sigma (\tau_2, \tau_3 ) &=& -\delta( \tau_1 - \tau_3 ) \nn
\Sigma ( \tau_1, \tau_2 ) &=& - J^2 \left[ G (\tau_1, \tau_2) \right]^2 G (\tau_2, \tau_1) \label{eqint}
\eea
A crucial property of these equations is that they are invariant under the time re-parameterization $\tau \rightarrow \sigma$, under which
\bea
\tau &=& f (\sigma) \nn
G(\tau_1 , \tau_2) &=& \left[ f' (\sigma_1) f' (\sigma_2) \right]^{-1/4} \frac{ g (\sigma_1)}{g (\sigma_2)} \, G(\sigma_1, \sigma_2) \nn
\Sigma (\tau_1 , \tau_2) &=& \left[ f' (\sigma_1) f' (\sigma_2) \right]^{-3/4} \frac{ g (\sigma_1)}{g (\sigma_2)} \, \Sigma (\sigma_1, \sigma_2) 
\label{eq:conf}
\eea
where $f(\sigma)$ and $g(\sigma)$ are arbitrary functions, corresponding to {\em emergent\/} conformal and U(1) gauge invariances.
The conformal symmetry of the low energy Green's functions has been noted earlier \cite{PG98,SS10,SS10b,AK15}, and in the form in Eq.~(\ref{eq:conf}) by Kitaev \cite{AK15} (without the $g (\sigma)$ factors).
The gauge transformation $g(\sigma)$ is a real number in Euclidean time, but it becomes a conventional U(1) phase factor in 
Minkowski time. For the original model of SY \cite{SY92}, the gauge invariance was explicitly present in the underlying Hamiltonian.
In contrast, our Hamiltonian here in Eq.~(\ref{H}) is not gauge-invariant, and only has a global U(1) symmetry; nevertheless, a U(1)
gauge invariance emerges in the low energy theory.

Note that the $i \omega_n$ term in Eq.~(\ref{eq:SY}) breaks both the conformal and gauge invariances. Although the $i \omega_n$ can mostly 
be neglected in studying the scaling limit, it is important in selecting the proper low energy solution of Eq.~(\ref{eqint}) from the highly degenerate possibilities allowed by
Eq.~(\ref{eq:conf}).

\subsection{Low energy Green's function}

We now show that the fermion Green's function in Eq.~(\ref{GR}) follows directly from the conformal and gauge invariances in Eq.~(\ref{eq:conf}),
when combined with constraints from analyticity and unitarity.
This Green's function was obtained earlier \cite{SY92,PG98,PGKS97,GPS00} by explicit solution of the integral equation
in Eq.~(\ref{eqint}) (and also, as discussed in Section~\ref{sec:holo}, by the solution \cite{Faulkner09,Faulkner11} of a Dirac equation on a thermal AdS$_2$ background with
a non-zero electric field). Given our reliance on conformal and gauge invariances, the computations below are straightforwardly generalized
to other operators with different values of $q$.

At the Matsubara frequencies, the Green's function is defined by
\beq
G(i \omega_n) = \int_0^{1/T} d \tau e^{i \omega_n \tau} G(\tau),
\eeq
and this is continued to all complex frequencies $z$ via
the spectral representation
\beq
G(z) = \int_{-\infty}^{\infty} \frac{d \Omega}{\pi} \frac{\rho (\Omega)}{z - \Omega}. \label{spec}
\eeq
The spectral density $\rho (\Omega ) > 0$ for all real $\Omega$ and $T$. The retarded Green's function is $G^R (\omega) = G(\omega + i \eta)$ with $\eta$ a positive infinitesimal, while the advanced Green's function is $G^A (\omega) = G(\omega - i \eta)$.

At $T=0$, given the scale invariance implicit in Eq.~(\ref{eq:conf}), we expect $G(z)$ to be a power-law of $z$. More precisely, 
Eq.~(\ref{eq:conf}) implies
\beq
G(z) = C \frac{e^{- i (\pi/4 +  \theta)}}{\sqrt{z}} \quad , \quad \mbox{Im} (z) > 0,~ |z| \ll J,~ T=0 . \label{Gz}
\eeq
Positivity of $\rho(\Omega)$ now implies $C>0$ and $-\pi/4 < \theta < \pi/4$.
An inverse Fourier transform yields
\bea 
G(\tau) = \left\{
\begin{array}{ccc}
\displaystyle - \frac{C \sin (\pi/4 + \theta)}{\sqrt{\pi \tau}} &,& \tau \gg 1/J,~T=0 \\[1em] 
\displaystyle \frac{C \cos (\pi/4 + \theta)}{\sqrt{-\pi \tau}} &,& -\tau \gg 1/J,~T=0.
\end{array} \right. \label{Gtau}
\eea

We obtain the non-zero temperature solution by choosing the conformal map in Eq.~(\ref{eq:conf}) as
\beq
\tau = \frac{1}{\pi T} \tan ( \pi T \sigma )
\eeq
where $\sigma$ is the periodic imaginary time co-ordinate with period $1/T$.
Applying this map to Eq.~(\ref{Gtau}) we obtain
\bea 
G(\sigma) = \left\{
\begin{array}{ccc}
\displaystyle - C g(\sigma) \sin (\pi/4 + \theta) \left( \frac{T}{\sin (\pi T \sigma)} \right)^{1/2} &,& \displaystyle 0 < \sigma < \frac{1}{T} \\[1em] 
\displaystyle  C g(\sigma) \cos (\pi/4 + \theta) \left( \frac{T}{\sin (-\pi T \sigma)} \right)^{1/2} &,& \displaystyle 0 < -\sigma < \frac{1}{T}.
\end{array} \right. \label{Gsigma}
\eea
The function $g (\sigma)$ is so far undetermined apart from a normalization choice $g(0)=1$.
We can now determine $g (\sigma) $ by imposing the KMS condition
\beq
G (\sigma + 1/T) = - G(\sigma)
\eeq
which implies
\beq
g(\sigma) = \tan(\pi/4 + \theta) g(\sigma + 1/T).
\eeq
The solution is clearly 
\beq
g(\sigma) = e^{-2 \pi \mathcal{E} T \sigma} 
\eeq
where the new parameter $\mathcal{E}$ and the angle $\theta$ are related as in Eq.~(\ref{theta}) for $\Delta=1/4$ and $q=1$.
The final expression determining $G(\sigma)$ is 
\beq
G(\sigma) =  - C \frac{e^{-2 \pi \mathcal{E} T \sigma}}{\sqrt{1 + e^{-4 \pi \mathcal{E}}}} \left( \frac{T}{\sin (\pi T \sigma)} \right)^{1/2} 
\quad,\quad 0 < \sigma < \frac{1}{T}, \label{Gsigma2}
\eeq
and this can be extended to all real $\sigma$ using the KMS condition. The result in Eq.~(\ref{GR}) now follows from a Fourier transform.

For other fermionic operators with general charge $q$ and scaling dimension $\Delta$, 
the above arguments show that the $\sigma$ dependence of 
Eq.~(\ref{Gsigma2}) will be replaced by
\beq
G(\sigma)  \sim - e^{-2 \pi q \mathcal{E} T \sigma} \left( \frac{T}{\sin (\pi T \sigma)} \right)^{2\Delta} \quad,\quad 0 < \sigma < \frac{1}{T}, \label{Gsigma3}
\eeq
and its Fourier transform will have the frequency shift
\beq
\omega_{\mathcal{S}} = 2 \pi q T \mathcal{E}. \label{omegaE}
\eeq
The $T \rightarrow 0$ limit of Eq.~(\ref{Gsigma3}) leads to Eq.~(\ref{twist}).

The above analysis can be easily repeated for bosonic operators of charge $q$ and scaling dimension $\Delta$. 
The result in Eq.~(\ref{GR}) continues to apply, while Eq.~(\ref{theta}) is modified to
\beq
e^{2 \pi q \mathcal{E}} = -\frac{\sin (\pi \Delta + \theta)}{\sin(\pi \Delta - \theta)}. \label{thetab}
\eeq
The constraint on the allowed values of $\theta $ is now $\pi \Delta < \theta < \pi(1 - \Delta)$.

The constants $C$ and $\theta$ (or $\mathcal{E}$) appearing in Eq.~(\ref{GR}) can also be determined exactly for the microscopic model
in Eq.~(\ref{H}), as reviewed in Appendix~\ref{app:nonuniv};
however, their values depend upon the specific UV completion used here, and do not apply to the
holographic model of Section~\ref{sec:holo}. In particular the `equation of state' for $\mathcal{Q}$ as a function of $\mathcal{E}$ 
is in Eq.~(\ref{QE}).

\subsection{Entropy}
\label{sec:entropy}

To complete our results for the SY state, 
we need to establish the connection in Eq.~(\ref{ws2}) between $\omega_{\mathcal{S}}$ and the zero temperature
entropy density $\mathcal{S}$. This is connection is the focus of our work, and it also relies on the conformal and 
gauge invariances in Eq.~(\ref{eq:conf}). However, in addition, we need information on the UV complete nature
of the fermion model, and in particular, the fact that the short-time behavior of the fermion Green's function is 
determined by canonical fermions obeying the anti-commutation relations in Eq.~(\ref{anti}).

The computation of the entropy follows Refs.~\onlinecite{PGKS97,GPS00}, and relies on the thermodynamic Maxwell relation
\beq
\left( \frac{\partial \mathcal{S}}{\partial \mathcal{Q}} \right)_{T} = -  \left( \frac{\partial \mu}{\partial T} \right)_{\mathcal{Q}}.
\label{maxwell}
\eeq
In the $T \rightarrow 0$ limit, Parcollet {\em et al.} \cite{PGKS97} (Section VI.A.2) show that the right-hand-side of Eq.~(\ref{maxwell}) can be evaluated using the imaginary time Green's function, and we review their computation here. 
Their argument requires not only the scaling behavior of the Green's function at times
$\tau \gg 1/J$ given in Eq.~(\ref{Gsigma3}), but also the short time behavior which is beyond the conformal regime.
First, we observe from Eq.~(\ref{eq:SY}) the large frequency behavior
\beq
G(i \omega_n) = \frac{1}{i \omega_n} - \frac{\mu}{(i \omega_n)^2} + \ldots
\eeq
which implies, from Eq.~(\ref{spec}),
\beq
\mu = -\int_{-\infty}^{\infty} \frac{d \Omega}{\pi} \, \Omega \rho (\Omega),
\eeq
which makes it evident that $\mu$ depends only upon the particle-hole asymmetric part of the spectral density.
Next, we can relate the $\Omega$ integrals to the derivative of the imaginary time correlator
\beq
\mu = - \partial_\tau G (\tau = 0^+) - \partial_\tau G(\tau = (1/T)^-). \label{mutau}
\eeq
Making the analogy to Eq.~(\ref{Gsigma2}), we pull out an explicitly particle-hole asymmetric part of $G(\tau)$ by defining
\beq
G(\tau) \equiv e^{-2 \pi \mathcal{E} T \tau} g(\tau) \quad,\quad 0 < \sigma < \frac{1}{T}. \label{gtaux}
\eeq
Note that $\mathcal{E}$ was introduced as a parameter in Eq.~(\ref{Gsigma2}), and then
appears in Eq.~(\ref{GR}) via Eq.~(\ref{omegaE}).
(It might appear that we can absorb the chemical potential, $\mu$, in the Hamiltonian by a temporal gauge transformation, and that such
a transformation combined with Eq.~(\ref{gtaux}) implies $\mu = - 2 \pi \mathcal{E} T$ and so yields $\partial \mu /\partial T$; however this argument is flawed because $\mu$ also includes a $\mathcal{Q}$-dependent piece at $T=0$, which must also be accounted for in
the gauge transformation. The non-zero $\mathcal{Q}$ ground state is not invariant under gauge transformations.)
We proceed in our computation of $\mu$ by inserting Eq.~(\ref{gtaux}) into Eq.~(\ref{mutau}) to obtain
\beq
\mu = 2 \pi \mathcal{E} T \left[ G (\tau = 0^+) + G(\tau = (1/T)^-) \right] - \partial_\tau g (\tau = 0^+)  - e^{-2 \pi \mathcal{E}} 
\partial_\tau g(\tau = (1/T)^-) \label{mu1}
\eeq
For the term in the first square brackets, we have
\beq
G (\tau = 0^+) + G(\tau = (1/T)^-) = G(\tau=0^{+}) - G(\tau=0^{-1}) = -1,
\eeq
which follows from the KMS condition and the fermion anti-commutation relation in Eq.~(\ref{anti}); also, this is related to the high frequency behavior $G(|z| \rightarrow \infty) = 1/z$. 
Writing the second term in Eq.~(\ref{mu1}) in terms of a spectral density $\rho_g (\Omega)$ for $g (\tau)$, we obtain
\beq
\mu = - 2 \pi \mathcal{E} T - \int_{-\infty}^{\infty} \frac{d\Omega}{\pi} \frac{\Omega \left[ \rho_g (\Omega) - e^{-2 \pi \mathcal{E}} \rho_g (-\Omega) \right]}{1 + e^{-\Omega/T}};
\eeq
(we note that there is a sign error on the right-hand-side of Eq. (65) in Ref.~\onlinecite{PGKS97}, and $-\rho_g(-\Omega)$ should be $\rho_g (-\Omega)$).
At this point, Ref.~\onlinecite{PGKS97} argues that at low $T$ and fixed $\mathcal{Q}$, $\rho_g$ must be particle hole symmetric with $\rho_g (\Omega) = \rho_g (- \Omega)$, and that the $T$ dependent part of the integral above scales as $T^{3/2}$.
We therefore have 
\beq
\left( \frac{\partial \mu}{\partial T} \right)_{\mathcal{Q}} = - 2 \pi \mathcal{E} \quad , \quad T \rightarrow 0, \label{muT1}
\eeq
and then the Maxwell relation in Eq.~(\ref{maxwell}) leads to Eq.~(\ref{SQE1b}).

Using the relationship between $\mathcal{Q}$ and $\mathcal{E}$ specified in Appendix~\ref{app:nonuniv} in Eq.~(\ref{QE}), and the limiting value $\mathcal{S} = 0$
in the empty state $\mathcal{Q}=0$, we can integrate Eq.~(\ref{SQE1b}) to obtain the full zero temperature entropy \cite{GPS00}.

\section{Charged black holes}
\label{sec:holo}

This section (apart from Section~\ref{blackhole}) 
mainly recalls the results of Faulkner {\em et al.\/} \cite{Faulkner09,Faulkner11} on planar, charged black holes in AdS$_{d+2}$,
and makes the correspondence with the properties of the SY state.  We will also largely follow their notation, apart from the change $d \rightarrow d+1$
required by our definition of $d$ as the spatial dimension (instead of the spacetime dimension). 
The case of spherical black holes in global
AdS is more complicated and is considered in Appendix~\ref{app:sphere}; it has a more complex equation of state, but also 
obeys all results claimed in 
Section~\ref{sec:intro}. The discussion in the latter part of Section~\ref{blackhole} shows how the needed features of Faulkner {\em et al.}
can be obtained in a more general class of black hole solutions.

We consider the Einstein-Maxwell theory of a metric $g$ and a U(1) gauge flux $F = dA$ with action
\beq
S = \frac{1}{2 \kappa^2} \int d^{d+2} x \sqrt{-g} \left[ \mathcal{R} + \frac{d(d+1)}{R^2} - \frac{R^2}{g_F^2} F^2 \right], \label{EM}
\eeq
where $\kappa^2 = 8 \pi G_N$, $\mathcal{R}$ is the Ricci scalar, 
$R$ is the radius of AdS$_{d+2}$, and $g_F$ is a dimensionless gauge coupling constant.
The equations of motion of this action have the solution \cite{Romans92,Myers99}
\beq
ds^2 = \frac{r^2}{R^2} \left( -f dt^2 + d \vec{x}^2 \right) + \frac{R^2}{r^2} \frac{dr^2}{f} \label{metric1}
\eeq
with
\bea
f &=& 1 + \frac{\Theta^2}{r^{2d}} - \left(r_0^{d+1} + \frac{\Theta^2}{r_0^{d-1}} \right) \frac{1}{r^{d+1}} \nn
A &=& \mu \left( 1 - \frac{r_0^{d-1}}{r^{d-1}} \right) dt \label{fA}
\eea
This solution is expressed in terms of three 
parameters $\Theta$, $r_0$, and $\mu$; these parameters are determined by the charge density, $\mathcal{Q}$, and temperature, $T$, 
of the boundary theory via the relations
\bea
\mu = \frac{g_F \Theta}{c_d R^2 r_0^{d-1}} \quad &,& \quad
\mathcal{Q} = \frac{2 (d-1)}{c_d} \frac{\Theta}{\kappa^2 R^d g_F} \nn
T = \frac{(d+1) r_0}{4 \pi R^2} \left( 1 - \frac{(d-1) \Theta^2}{(d+1) r_0^{2d}} \right) \quad &,& \quad c_d = \sqrt{\frac{2(d-1)}{d}}.
\label{liueqs}
\eea
The Bekenstein-Hawking entropy density \cite{JDB73,SWH75} of this solution is
\beq
\mathcal{S}_{\rm BH} = \frac{2 \pi}{\kappa^2} \left(\frac{r_0}{R} \right)^d . \label{Ah1}
\eeq

We turn next to the holographic implications of this solution at low energy \cite{Faulkner09,Faulkner11,TFJP11},
which is controlled by the near-horizon geometry. At $T=0$, the horizon is at $r= \left[ \Theta^2 (d-1)/(d+1) \right]^{1/(2d)}$, and so we introduce
the co-ordinate $\zeta$ by
\beq
r - \left[ \Theta^2 (d-1)/(d+1) \right]^{1/(2d)} = \frac{1}{\zeta}; \label{rzeta}
\eeq
we approach the horizon as $\zeta \rightarrow \infty$ (see Fig.~\ref{fig:summary}).
In terms of $\zeta$, the near horizon geometry at $T=0$ is
\beq
ds^2 = R_2^2 \frac{(-dt^2 + d\zeta^2)}{\zeta^2} + \frac{\left[ \Theta^2 (d-1)/(d+1) \right]^{1/d}}{R^2}  d \vec{x}^2. \label{factor}
\eeq
The geometry has factorized to AdS$_2 \times \mathbb{R}^d$, where the AdS$_2$ radius is given by
\beq
R_2 = \frac{R}{\sqrt{d(d+1)}}. \label{planarR2}
\eeq
In the same low energy limit
the gauge field is (see Fig.~\ref{fig:summary})
\beq
A = \frac{\mathcal{E}}{\zeta} dt. \label{Azeta}
\eeq
which determines the strength of the AdS$_2$ electric field in terms of the dimensionless parameter $\mathcal{E}$. 
Notice that the value of $\mathcal{E}$ in Eq.~(\ref{Azeta}) is invariant 
under any rescaling of the co-ordinates which
preserves the $(-dt^2 + d \zeta^2)/\zeta^2$ structure of the AdS$_2$ metric. From the present near-horizon computation we find
the value
\beq
\mathcal{E} = 
\frac{g_F \, \mbox{sgn}(\mathcal{Q})}{\sqrt{2d (d+1)}}. \label{SQE4}
\eeq
Eq.~(\ref{SQE4}) is the `equation of state' connecting $\mathcal{Q}$ to $\mathcal{E}$, and the analogous expression for the fermion model is in Eq.~(\ref{QE}), and for the spherical black hole is in Eq.~(\ref{s4}); the non-analytic $\mathcal{Q}$ dependence in Eq.~(\ref{SQE4})
becomes analytic for the spherical black hole in Appendix~\ref{app:sphere}.
We recall that $g_F$ is a dimensionless coupling, and so $\mathcal{E}$ is also dimensionless, and depends only upon $g_F$ and $d$;
in particular, $\mathcal{E}$ is independent of $\kappa^2$, and so remains of order unity in the matrix large $M$ limit of 
holography \cite{Maldacena97}, as noted in Section~\ref{sec:intro}.

We also take the $T=0$ limit of Eq.~(\ref{Ah1}) from Eq.~(\ref{liueqs}), and find 
\beq
\mathcal{S}_{\rm BH} = \frac{2 \pi g_F |\mathcal{Q}|}{\sqrt{2d(d+1)}} \quad, \quad T \rightarrow 0 . \label{Ah2}
\eeq
Comparing Eqs.~(\ref{SQE4}) and (\ref{Ah2}), we find that Eq.~(\ref{SQE1b}) is indeed obeyed. 
Note that for the present case of a planar black hole, we can combine Eqs.~(\ref{SQE4}) and (\ref{Ah2}) into the simple
relationship \cite{Faulkner09}
\beq
\mathcal{S}_{\rm BH} = 2 \pi \mathcal{Q} \mathcal{E}. \label{BH1}
\eeq
Eq.~(\ref{BH1}) does not hold for a spherical black hole;
but the more fundamental relation for $\partial \mathcal{S}_{\rm BH}/\partial \mathcal{Q}$ in Eq.~(\ref{SQE1b}) does hold,
and is verified in Appendix~\ref{app:sphere}, which also derives  
the different `equation of state' relating $\mathcal{E}$ and $\mathcal{Q}$ for a spherical black hole.

\subsection{Fermion correlations}

To confirm the link to the fermion model, we need to show that the $\mathcal{E}$ obtained above in Eq.~(\ref{Azeta}) is the
same as the $\mathcal{E}$ (or the related $\omega_\mathcal{S}$ via Eq.~(\ref{SQE1a})) appearing as the spectral asymmetry parameter in
the response functions in Eqs.~(\ref{GR}) and (\ref{twist}) (see Fig.~\ref{fig:summary}).
For this, we need the Green's function of matter fields moving on a thermal AdS$_2$ metric. The finite temperature generalization of the
AdS$_2$ factor in Eq.~(\ref{factor}) is \cite{Faulkner09,Faulkner11}
\beq
ds^2 = \frac{R_2^2}{\zeta^2} \left[ - \left(  1- {\zeta^2}/{\zeta_0^2} \right) dt^2 + \frac{d \zeta^2}{(1 - \zeta^2/\zeta_0^2)} \right], \label{ads2}
\eeq
and that of the gauge field is
\beq
A = \mathcal{E} \left(\frac{1}{\zeta} - \frac{1}{\zeta_0} \right) dt \label{aads2}
\eeq
where 
\beq
T = \frac{1}{2 \pi \zeta_0}. \label{Tzeta}
\eeq 
The action of a fermionic spinor, $\psi$, of charge $q$ moving in the backgrounds of Eqs.~(\ref{ads2}) and (\ref{aads2}) is
\beq
S = i \int d^2 x \sqrt{-g} \left( \overline{\psi} \Gamma^\alpha D_\alpha \psi - m \overline{\psi} \psi \right) \label{Sads2}
\eeq
where $m$ is a bulk fermion mass, $\Gamma^\alpha$ are the Dirac Gamma matrices, and $D_\alpha$ is a covariant derivative with charge $q$. 
The correlator of $\psi$ in this thermal AdS$_2$ \cite{Iqbal09} plus electric field background has been computed in some detail 
by Faulkner {\em et al.\/} \cite{Faulkner09,Faulkner11}, and their result was already displayed in Eq.~(\ref{GR})
in our notation. This computation shows that $\mathcal{E} = \omega_{\mathcal{S}}/(2 \pi q T)$ (Eq.~(\ref{SQE1a})) is indeed the same parameter appearing in 
Eqs.~(\ref{Azeta}) and (\ref{aads2}).
In this AdS$_2$ computation,
the scaling dimension $\Delta$ is related to the bulk spinor mass by
\beq
\Delta = \frac{1}{2} - \sqrt{m^2 R_2^2 - q^2 \mathcal{E}^2}. \label{nu}
\eeq
 
\subsection{Black hole thermodynamics}
\label{blackhole}
 
We close this section by noting a significant property of the above solution of classical general relativity at all $T$ and $\mathcal{Q}$.
From the laws of black holes thermodynamics \cite{Bardeen73}, we deduce that the horizon area and the chemical potential must obey
a thermodynamic Maxwell relation
\beq
 \left( \frac{\partial \mathcal{S}_{\rm BH}}{\partial \mathcal{Q}} \right)_{T} = -  \left( \frac{\partial \mu}{\partial T} \right)_{\mathcal{Q}}, \label{maxwell2}
\eeq
which is the analog of that in the fermion model computation in Eq.~(\ref{maxwell}).
And indeed we do find from Eqs.~(\ref{liueqs}) and (\ref{Ah1}) that Eq.~(\ref{maxwell2}) is obeyed with
\beq
\left( \frac{\partial \mu}{\partial T} \right)_{\mathcal{Q}}
= -\frac{4 \pi (d-1) g_F \Theta r_0^d}{
c_d  (d+1) r_0^{2d} + c_d (d-1)(2d-1) \Theta^2}. \label{maxwell3}
\eeq

In determining the value of $(\partial \mu /\partial T)_\mathcal{Q}$ as $T \rightarrow 0$,
rather than explicitly evaluating Eq.~(\ref{maxwell3}), it is instructive to use a more general argument which
does not use the explicit form of the solution in Eqs.~(\ref{fA}) and (\ref{liueqs}). 
From the original action in Eq.~(\ref{EM}) and the metric in Eq.~(\ref{metric1}), Gauss's law for the scalar potential in the bulk is
\beq
\frac{2 R^2}{\kappa^2 g_F^2} \frac{d}{dr}\left( \frac{r^d}{R^d} \frac{d A_t}{d r} \right) = 0 , \label{gauss1}
\eeq
and the constant of integration is the boundary charge density
\beq
\frac{2 R^2}{\kappa^2 g_F^2} \left( \frac{r^d}{R^d} \frac{d A_t}{d r} \right) = \mathcal{Q}. \label{gauss2}
\eeq
We can write the solution of Eq.~(\ref{gauss2}) as
\beq
A_t (r) = \mu (T) - \left( \frac{R^{d-2} \kappa^2 g_F^2}{2 (d-1)} \right) \frac{\mathcal{Q}}{r^{d-1}}, \label{At}
\eeq
where the $r$-dependent term in Eq.~(\ref{At}) is independent of $T$ at fixed $\mathcal{Q}$, and
the chemical potential $\mu$ equals $A_t (r \rightarrow \infty)$ when we choose $A_t=0$ on the horizon.
Now we transform to the near-horizon AdS$_2$ geometry by making a $T$-independent change of variables from $r$ to $\zeta$
as in Eq.~(\ref{rzeta}), $r = r_{\ast} + 1/\zeta$, where $r=r_{\ast}$ is the position of the horizon at $T=0$, but we won't need
the actual value of $r_{\ast}$. Then Eq.~(\ref{At}) implies that, as $\zeta \rightarrow \infty$,  
the near-horizon scalar potential must of the form in Eq.~(\ref{aads2}), where now we
define
$\zeta=\zeta_0$ as the position of the horizon at non-zero $T$, where $\mathcal{E}$ is a parameter independent of $T$, and
\beq
\left( \frac{\partial \mu}{\partial T} \right)_{\mathcal{Q}} =  \mathcal{E} \frac{\partial}{\partial T} \left( - \frac{1}{\zeta_0} \right)_{\mathcal{Q}}.
\eeq
The $T$-dependence of $\zeta_0$ in Eq.~(\ref{Tzeta}) follows from the conformal mapping between the $T=0$ AdS$_2$ metric in Eq.~(\ref{factor}) and $T>0$ metric in Eq.~(\ref{aads2}) \cite{Faulkner11}. So we find by this general argument that
\beq
\left( \frac{\partial \mu}{\partial T} \right)_{\mathcal{Q}} = - 2 \pi \mathcal{E} \quad, \quad T \rightarrow 0, \label{muT2}
\eeq
which is the same as the fermion model result in Eq.~(\ref{muT1}). It can be verified that Eq.~(\ref{muT2})
holds also in the spherical geometry of Appendix~\ref{app:sphere}.
Combining Eq.~(\ref{muT2}) with Eq.~(\ref{maxwell2}), we obtain Eq.~(\ref{SQE1b}), which is a special case 
of results obtained from the Wald formalism \cite{Sen05,Sen08,Wald93,Myers93,Wald94,Myers94,Shahin13}.

We note that the above derivation of Eq.~(\ref{muT2}) relied only on Gauss's Law and the conformal invariance of the
AdS$_2$ near-horizon geometry: this implies that such results hold for a wide class of black hole solutions \cite{Sen05,Sen08,Wald93,Myers93,Wald94,Myers94,Shahin13}.
 
\section{Discussion}
\label{sec:conc}

In our discussion of the SY state of the infinite-range fermion model in Eq.~(\ref{H}), we noted that the fermion Green's function was almost completely determined by the emergent conformal and gauge invariances in Eq.~(\ref{eq:conf}). These conformal and gauge invariances also fairly uniquely 
determine the holographic theory of matter moving in curved space in the presence of an electric field. So, with the benefit of hindsight, 
we can understand the equivalence of the fermion Green's functions obtained in Sections~\ref{sec:inf} and~\ref{sec:holo}.

However, we have gone beyond the identification of Green's functions, and also shown that the zero temperature entropy of the SY state
can be mapped onto that of the AdS$_2$ theory (see Fig.~\ref{fig:summary}).  
Specifically, we chose an appropriate combination of observables in 
Eqs.~(\ref{ws1},\ref{ws2})
to allow us to generally define a common frequency $\omega_{\mathcal{S}}$, and we showed that this frequency was related to
precisely the same derivative of the entropy in both the SY state and in charged black holes (where the entropy was the Bekenstein-Hawking entropy). In both cases, establishing this relationship required an analysis
of the details of the model, and it did not follow from general symmetry arguments alone. 
In particular, for the SY state, the entropy computation required careful treatment of the manner in which the emergent gauge and 
conformal invariances, present at low energies, were broken by the on-site canonical fermions, present at high energies.

This common relationship between $\omega_{\mathcal{S}}$ and the entropy 
indicates an equivalence between the low-energy degrees of freedom of the two theories in Sections~\ref{sec:inf} and~\ref{sec:holo}, and
strongly supports the existence of gravity dual of the SY state with a AdS$_2$ horizon.
The present results also imply the $c_i$ fermion, with $q=1$, of the theory in Eq.~(\ref{H}) is holographically dual to the $\psi$ fermion, with $q=1$, \cite{Faulkner09,TFJP11}
of Eq.~(\ref{Sads2}). As the microscopic 
$c_i$ fermion carries all of the $\mathcal{Q}$ charge of the theory in Eq.~(\ref{H}), we expect that $\psi$ also
carries a non-negligible fraction of the charge (in the large $N$ limit) behind the AdS$_2$ horizon. Both models likely also have higher
dimension operators, but these have not been analyzed so far (see however Ref.~\onlinecite{AK15}).

Note that the above discussion refers to the near-horizon AdS$_2$ geometry. The larger Reissner-Nordstr\"om-AdS solution is to
be regarded here as a convenient (and non-universal) embedding space which provides a UV regulation of the gravitational theory.
With such an embedding, we are able compute well-defined values for $\mathcal{S}$ and $\mathcal{Q}$.
Presumably other gravitational UV embeddings, will have different `equations of state' between $\mathcal{E}$ and $\mathcal{Q}$,
but the will nevertheless obey the fundamental relation in Eq.~(\ref{SQE1b}) provided they contain a AdS$_2$ horizon.
We explicitly tested the independence on the UV embedding in Appendix~\ref{app:sphere}
by comparing the cases of planar and spherical black holes.

The above identification between the $c_i$ and $\psi$ fermions differs from that made previously by the author in Refs.~\onlinecite{SS10,SS10b}.
There, $\psi$ was argued to be dual to a higher dimension composite fermion operator of the original model of SY \cite{SY92}. 
This previous identification was based
upon the requirement that local bulk operators must be dual to gauge-invariant operators on the boundary, and the original model \cite{SY92} had
a microscopic gauge invariance which did not allow the choice of $c_i$ as dual to $\psi$. 
However, in the present model in Eq.~(\ref{H}), 
there is no microscopic gauge invariance, and so we are free to use $c_i$ as the dual of the bulk $\psi$ field. It turns out that the low energy
boundary theory for $c_i$ does have a gauge invariance (as in Eq.~(\ref{eq:conf})), but this is an emergent gauge invariance which is broken
by UV terms needed to regularize the theory.
The present situation is analogous to the theory of the Ising-nematic quantum critical point in metals, where the regularized model for the electrons
is not gauge-invariant, but the low energy theory defined on two Fermi surface patches does have an emergent gauge structure \cite{MS10,MMLS10}. 
And the present situation is different from that in the `slave particle' theories of condensed matter, where the gauge structure emerges
from fractionalizing particles into partons, which influenced the reasoning of Refs.~\onlinecite{SS10,SS10b}.
Instead the same particle can be gauge-invariant in the underlying theory, and acquire an emergent gauge charge in the low energy theory.
There is some similarity between this interpretation and ideas in Ref.~\onlinecite{Gubser11}.

Finally, we note recent work \cite{Shenker13,Shenker15,AK15} on `a bound on chaos' which also related characteristic times of the real-time dynamics of strongly-coupled quantum systems to thermodynamics, $\hbar$, and black hole horizons.

\section*{Acknowledgments} 
I thank T.~Banks, A.~Dabholkar, Wenbo Fu, S.~Hartnoll, A.~Kitaev, Hong Liu, J.~McGreevy, R.~Myers, A.~Sen, A.~Strominger, and W.~Witczak-Krempa for valuable discussions,
and especially A.~Georges and O.~Parcollet for inspiring discussions on these topics over many years.
This research was supported by the NSF under Grant DMR-1360789, and also partially by the Templeton Foundation. 
The research at KITP Santa Barbara was supported by 
the Simons Foundation and NSF Grant PHY11-25915.
Research at Perimeter Institute is supported by the Government of Canada through Industry Canada 
and by the Province of Ontario through the Ministry of Research and Innovation.   
 
\appendix

\section{Non-universal constants of the fermion model}
\label{app:nonuniv}

We compute the constants $C$ and $\theta$ (or $\mathcal{E}$) appearing in Eq.~(\ref{GR}) for the microscopic model
in Eq.~(\ref{H}). The results of this appendix do not apply to the
holographic model of Section~\ref{sec:holo}.

We can compute the self-energy from Eq.~(\ref{Gtau}) and the second equation in Eq.~(\ref{eq:SY}) 
\bea 
\Sigma (\tau) = \left\{
\begin{array}{ccc}
\displaystyle - \frac{C^3 J^2 \cos(2 \theta) \sin (\pi/4 + \theta) }{2 (\pi \tau)^{3/2}} &,& \tau \gg J,~T=0 \\[1em] 
\displaystyle \frac{C^3 J^2 \cos(2 \theta) \cos (\pi/4 + \theta) }{2 (-\pi \tau)^{3/2}}  &,& -\tau \gg J,~T=0
\end{array} \right. .\label{Stau}
\eea
A Fourier transform now leads to 
\beq
\Sigma (z) = - \frac{J^2 C^3 \cos (2 \theta)}{\pi} e^{ i (\pi/4 + \theta)} \sqrt{z} \quad , \quad \mbox{Im} (z) > 0,~ |z| \ll J,~T=0 . \label{Sz}
\eeq
We now see that Eqs.~(\ref{Gz}) and (\ref{Sz}) are consistent with the first equation in Eq.~(\ref{eq:SY}), provided we choose the value of $C$ to be 
\beq
C = \left( \frac{\pi}{J^2 \cos (2 \theta) }\right)^{1/4}.
\eeq

Finally, the value of $\theta$ can be related to the density $\mathcal{Q}$ by a computation which parallels the
Luttinger-Ward analysis \cite{LW60} for a Fermi liquid. The present model has no spatial structure, and so no possibility of a Fermi surface.
However, if we apply the steps of the Luttinger-Ward proof of the volume enclosed by the Fermi surface, we find an expression relating
density $\mathcal{Q}$ to the spectral asymmetry angle $\theta$. In other words, $\theta$ plays a role similar to the Fermi wavevector
in a Fermi liquid. And the relationship between $\mathcal{Q}$ and $\theta$ is \cite{GPS00}
\beq
\mathcal{Q} = \frac{1}{2} - \frac{\theta}{\pi} - \frac{\sin (2 \theta)}{4}. \label{lutt1}
\eeq
Note that the constraint $-\pi/4<\theta<\pi/4$ implies that $0 < \mathcal{Q} < 1$, as expected.
In terms of $\mathcal{E}$, this relationship is
\beq
\mathcal{Q} = \frac{1}{4} (3 - \tanh (2 \pi \mathcal{E})) - \frac{1}{\pi} \tan^{-1} \left( e^{2 \pi \mathcal{E} } \right). \label{QE}
\eeq
The right-hand-side is a monotonically decreasing function of $\mathcal{E}$ which ranges between 1 and 0, as $\mathcal{E}$ increases from
$-\infty$ to $\infty$.

\section{Spherical black holes}
\label{app:sphere}

We consider the case of spherical black holes in global AdS, following the analysis of Ref.~\onlinecite{Myers99}.
For simplicity, we will limit ourselves to the $T=0$ case. 

Now we choose a solution of the Einstein-Maxwell equations of motion of Eq.~(\ref{EM}) with metric
\beq
ds^2 = - V(r) dt^2 + r^2 d \Omega_d^2 + \frac{dr^2}{V(r)} \label{s1}
\eeq
where $d \Omega_d^2$ is the metric of the $d$-sphere, and
\beq
V(r) = 1 + \frac{r^2}{R^2} + \frac{\Theta^2}{r^{2d-2}} - \frac{M}{r^{d-1}},
\eeq
has a zero at $r=r_0$ so that
\beq
M = r_0^{d-1} \left( 1 + \frac{r_0^2}{R^2} + \frac{\Theta^2}{r_0^{2d-2}} \right).
\eeq
The zero temperature case has \cite{Myers99}
\beq
\Theta^2 = \frac{r_0^{2d-2} \left[ (d-1)R^2 + (d+1) r_0^2 \right]}{(d-1)R^2}.
\eeq
In the near-horizon region, we introduce, as in Section~\ref{sec:holo}, the co-ordinate $\zeta$ via
\beq
r - r_0 = \frac{R_2^2}{\zeta},
\eeq
where Eq.~(\ref{planarR2}) is now replaced by
\beq
R_2 = \frac{R}{\sqrt{d(d+1) + (d-1)^2 R^2/r_0^2}},
\eeq
and the near-horizon metric becomes AdS$_2 \times$ S$_d$, with
\beq
ds^2 = R_2^2 \left[ \frac{-dt^2 + d \zeta^2}{\zeta^2} \right] + r_0^2 d \Omega_d^2.
\eeq

Turning to the gauge field sector, the charge density, $\mathcal{Q}$, and AdS$_2$ electric field parameter $\mathcal{E}$ in Eq.~(\ref{Azeta}) are
\bea
\mathcal{Q} &=& \frac{r_0^{d-1} \sqrt{2d \left[ (d-1)R^2 + (d+1)r_0^2 \right]}}{\kappa^2 g_F} \nn
\mathcal{E} &=& \frac{g_F r_0 \sqrt{2d \left[ (d-1)R^2 + (d+1)r_0^2 \right]}}{2 \left[ (d-1)^2 R^2 + d(d+1) r_0^2 \right]}. \label{s4}
\eea
The `equation  of state' obeyed by $\mathcal{E}$ and $\mathcal{Q}$ is obtained by eliminating $r_0$ between the equations
in Eq.~(\ref{s4}); this leads to a very lengthy expression which we shall not write out explicitly.

Using the Bekenstein-Hawking entropy density
\beq
\mathcal{S}_{\rm BH} = \frac{2 \pi}{\kappa^2} r_0^d, \label{s5}
\eeq
and 
\beq
\frac{\partial \mathcal{S}_{\rm BH}}{\partial \mathcal{Q}} = \frac{\partial \mathcal{S}_{\rm BH}/\partial r_0}{\partial \mathcal{Q}/\partial r_0},\label{s6}
\eeq
and evaluating the derivatives via Eq.~(\ref{s4}), we can now verify that
Eq.~(\ref{SQE1b}) is indeed obeyed.
Note that $\mathcal{S}_{\rm BH} \neq
2 \pi \mathcal{Q} \mathcal{E}$ here, unlike Eq.~(\ref{BH1}) for the planar case. 

\bibliography{ads}

\begin{thebibliography}{42}%
\makeatletter
\providecommand \@ifxundefined [1]{%
 \@ifx{#1\undefined}
}%
\providecommand \@ifnum [1]{%
 \ifnum #1\expandafter \@firstoftwo
 \else \expandafter \@secondoftwo
 \fi
}%
\providecommand \@ifx [1]{%
 \ifx #1\expandafter \@firstoftwo
 \else \expandafter \@secondoftwo
 \fi
}%
\providecommand \natexlab [1]{#1}%
\providecommand \enquote  [1]{``#1''}%
\providecommand \bibnamefont  [1]{#1}%
\providecommand \bibfnamefont [1]{#1}%
\providecommand \citenamefont [1]{#1}%
\providecommand \href@noop [0]{\@secondoftwo}%
\providecommand \href [0]{\begingroup \@sanitize@url \@href}%
\providecommand \@href[1]{\@@startlink{#1}\@@href}%
\providecommand \@@href[1]{\endgroup#1\@@endlink}%
\providecommand \@sanitize@url [0]{\catcode `\\12\catcode `\$12\catcode
  `\&12\catcode `\#12\catcode `\^12\catcode `\_12\catcode `\%12\relax}%
\providecommand \@@startlink[1]{}%
\providecommand \@@endlink[0]{}%
\providecommand \url  [0]{\begingroup\@sanitize@url \@url }%
\providecommand \@url [1]{\endgroup\@href {#1}{\urlprefix }}%
\providecommand \urlprefix  [0]{URL }%
\providecommand \Eprint [0]{\href }%
\providecommand \doibase [0]{http://dx.doi.org/}%
\providecommand \selectlanguage [0]{\@gobble}%
\providecommand \bibinfo  [0]{\@secondoftwo}%
\providecommand \bibfield  [0]{\@secondoftwo}%
\providecommand \translation [1]{[#1]}%
\providecommand \BibitemOpen [0]{}%
\providecommand \bibitemStop [0]{}%
\providecommand \bibitemNoStop [0]{.\EOS\space}%
\providecommand \EOS [0]{\spacefactor3000\relax}%
\providecommand \BibitemShut  [1]{\csname bibitem#1\endcsname}%
\let\auto@bib@innerbib\@empty
\bibitem [{\citenamefont {Maldacena}(1999)}]{Maldacena97}%
  \BibitemOpen
  \bibfield  {author} {\bibinfo {author} {\bibfnamefont {J.~M.}\ \bibnamefont
  {Maldacena}},\ }\bibfield  {title} {\enquote {\bibinfo {title} {{The Large N
  limit of superconformal field theories and supergravity}},}\ }\href {\doibase
  10.1023/A:1026654312961} {\bibfield  {journal} {\bibinfo  {journal} {Int. J.
  Theor. Phys.}\ }\textbf {\bibinfo {volume} {38}},\ \bibinfo {pages} {1113}
  (\bibinfo {year} {1999})},\ \Eprint {http://arxiv.org/abs/hep-th/9711200}
  {arXiv:hep-th/9711200 [hep-th]} \BibitemShut {NoStop}%
\bibitem [{\citenamefont {{Hartnoll}}(2011)}]{Hartnoll11}%
  \BibitemOpen
  \bibfield  {author} {\bibinfo {author} {\bibfnamefont {S.~A.}\ \bibnamefont
  {{Hartnoll}}},\ }\bibfield  {title} {\enquote {\bibinfo {title} {{Horizons,
  holography and condensed matter}},}\ }\href@noop {} {\bibfield  {journal}
  {\bibinfo  {journal} {ArXiv e-prints}\ } (\bibinfo {year} {2011})},\ \Eprint
  {http://arxiv.org/abs/1106.4324} {arXiv:1106.4324 [hep-th]} \BibitemShut
  {NoStop}%
\bibitem [{\citenamefont {Sachdev}(2012)}]{SSarcmp}%
  \BibitemOpen
  \bibfield  {author} {\bibinfo {author} {\bibfnamefont {S.}~\bibnamefont
  {Sachdev}},\ }\bibfield  {title} {\enquote {\bibinfo {title} {{What can
  gauge-gravity duality teach us about condensed matter physics?}}}\ }\href
  {\doibase 10.1146/annurev-conmatphys-020911-125141} {\bibfield  {journal}
  {\bibinfo  {journal} {Annual Review of Condensed Matter Physics}\ }\textbf
  {\bibinfo {volume} {3}},\ \bibinfo {pages} {9} (\bibinfo {year} {2012})},\
  \Eprint {http://arxiv.org/abs/1108.1197} {arXiv:1108.1197 [cond-mat.str-el]}
  \BibitemShut {NoStop}%
\bibitem [{\citenamefont {{Iqbal}}\ \emph {et~al.}(2012)\citenamefont
  {{Iqbal}}, \citenamefont {{Liu}},\ and\ \citenamefont {{Mezei}}}]{Iqbal11}%
  \BibitemOpen
  \bibfield  {author} {\bibinfo {author} {\bibfnamefont {N.}~\bibnamefont
  {{Iqbal}}}, \bibinfo {author} {\bibfnamefont {H.}~\bibnamefont {{Liu}}}, \
  and\ \bibinfo {author} {\bibfnamefont {M.}~\bibnamefont {{Mezei}}},\
  }\bibfield  {title} {\enquote {\bibinfo {title} {{Lectures on Holographic
  Non-Fermi Liquids and Quantum Phase Transitions}},}\ }in\ \href {\doibase
  10.1142/9789814350525_0013} {\emph {\bibinfo {booktitle} {String Theory and
  its Applications - TASI 2010, From meV to the Planck Scale}}},\ \bibinfo
  {editor} {edited by\ \bibinfo {editor} {\bibfnamefont {M.}~\bibnamefont
  {{Dine}}}, \bibinfo {editor} {\bibfnamefont {T.}~\bibnamefont {{Banks}}}, \
  and\ \bibinfo {editor} {\bibfnamefont {S.}~\bibnamefont {{Sachdev}}}}\
  (\bibinfo {year} {2012})\ pp.\ \bibinfo {pages} {707--815},\ \Eprint
  {http://arxiv.org/abs/1110.3814} {arXiv:1110.3814 [hep-th]} \BibitemShut
  {NoStop}%
\bibitem [{\citenamefont {{Sachdev}}\ and\ \citenamefont {{Ye}}(1993)}]{SY92}%
  \BibitemOpen
  \bibfield  {author} {\bibinfo {author} {\bibfnamefont {S.}~\bibnamefont
  {{Sachdev}}}\ and\ \bibinfo {author} {\bibfnamefont {J.}~\bibnamefont
  {{Ye}}},\ }\bibfield  {title} {\enquote {\bibinfo {title} {{Gapless
  spin-fluid ground state in a random quantum Heisenberg magnet}},}\ }\href
  {\doibase 10.1103/PhysRevLett.70.3339} {\bibfield  {journal} {\bibinfo
  {journal} {\prl}\ }\textbf {\bibinfo {volume} {70}},\ \bibinfo {pages} {3339}
  (\bibinfo {year} {1993})},\ \Eprint {http://arxiv.org/abs/cond-mat/9212030}
  {cond-mat/9212030} \BibitemShut {NoStop}%
\bibitem [{\citenamefont {{Parcollet}}\ \emph {et~al.}(1998)\citenamefont
  {{Parcollet}}, \citenamefont {{Georges}}, \citenamefont {{Kotliar}},\ and\
  \citenamefont {{Sengupta}}}]{PGKS97}%
  \BibitemOpen
  \bibfield  {author} {\bibinfo {author} {\bibfnamefont {O.}~\bibnamefont
  {{Parcollet}}}, \bibinfo {author} {\bibfnamefont {A.}~\bibnamefont
  {{Georges}}}, \bibinfo {author} {\bibfnamefont {G.}~\bibnamefont
  {{Kotliar}}}, \ and\ \bibinfo {author} {\bibfnamefont {A.}~\bibnamefont
  {{Sengupta}}},\ }\bibfield  {title} {\enquote {\bibinfo {title}
  {{Overscreened multichannel SU(N) Kondo model: Large-N solution and conformal
  field theory}},}\ }\href {\doibase 10.1103/PhysRevB.58.3794} {\bibfield
  {journal} {\bibinfo  {journal} {\prb}\ }\textbf {\bibinfo {volume} {58}},\
  \bibinfo {pages} {3794} (\bibinfo {year} {1998})},\ \Eprint
  {http://arxiv.org/abs/cond-mat/9711192} {cond-mat/9711192} \BibitemShut
  {NoStop}%
\bibitem [{\citenamefont {{Georges}}\ \emph {et~al.}(2001)\citenamefont
  {{Georges}}, \citenamefont {{Parcollet}},\ and\ \citenamefont
  {{Sachdev}}}]{GPS00}%
  \BibitemOpen
  \bibfield  {author} {\bibinfo {author} {\bibfnamefont {A.}~\bibnamefont
  {{Georges}}}, \bibinfo {author} {\bibfnamefont {O.}~\bibnamefont
  {{Parcollet}}}, \ and\ \bibinfo {author} {\bibfnamefont {S.}~\bibnamefont
  {{Sachdev}}},\ }\bibfield  {title} {\enquote {\bibinfo {title} {{Quantum
  fluctuations of a nearly critical Heisenberg spin glass}},}\ }\href {\doibase
  10.1103/PhysRevB.63.134406} {\bibfield  {journal} {\bibinfo  {journal}
  {\prb}\ }\textbf {\bibinfo {volume} {63}},\ \bibinfo {eid} {134406} (\bibinfo
  {year} {2001})},\ \Eprint {http://arxiv.org/abs/cond-mat/0009388}
  {cond-mat/0009388} \BibitemShut {NoStop}%
\bibitem [{\citenamefont {{Parcollet}}\ and\ \citenamefont
  {{Georges}}(1999)}]{PG98}%
  \BibitemOpen
  \bibfield  {author} {\bibinfo {author} {\bibfnamefont {O.}~\bibnamefont
  {{Parcollet}}}\ and\ \bibinfo {author} {\bibfnamefont {A.}~\bibnamefont
  {{Georges}}},\ }\bibfield  {title} {\enquote {\bibinfo {title}
  {{Non-Fermi-liquid regime of a doped Mott insulator}},}\ }\href {\doibase
  10.1103/PhysRevB.59.5341} {\bibfield  {journal} {\bibinfo  {journal} {\prb}\
  }\textbf {\bibinfo {volume} {59}},\ \bibinfo {pages} {5341} (\bibinfo {year}
  {1999})},\ \Eprint {http://arxiv.org/abs/cond-mat/9806119} {cond-mat/9806119}
  \BibitemShut {NoStop}%
\bibitem [{\citenamefont {{Damle}}\ and\ \citenamefont
  {{Sachdev}}(1997)}]{damless}%
  \BibitemOpen
  \bibfield  {author} {\bibinfo {author} {\bibfnamefont {K.}~\bibnamefont
  {{Damle}}}\ and\ \bibinfo {author} {\bibfnamefont {S.}~\bibnamefont
  {{Sachdev}}},\ }\bibfield  {title} {\enquote {\bibinfo {title}
  {{Nonzero-temperature transport near quantum critical points}},}\ }\href
  {\doibase 10.1103/PhysRevB.56.8714} {\bibfield  {journal} {\bibinfo
  {journal} {\prb}\ }\textbf {\bibinfo {volume} {56}},\ \bibinfo {pages} {8714}
  (\bibinfo {year} {1997})},\ \Eprint {http://arxiv.org/abs/cond-mat/9705206}
  {cond-mat/9705206} \BibitemShut {NoStop}%
\bibitem [{\citenamefont {Kovtun}\ \emph {et~al.}(2005)\citenamefont {Kovtun},
  \citenamefont {Son},\ and\ \citenamefont {Starinets}}]{kss}%
  \BibitemOpen
  \bibfield  {author} {\bibinfo {author} {\bibfnamefont {P.}~\bibnamefont
  {Kovtun}}, \bibinfo {author} {\bibfnamefont {D.~T.}\ \bibnamefont {Son}}, \
  and\ \bibinfo {author} {\bibfnamefont {A.~O.}\ \bibnamefont {Starinets}},\
  }\bibfield  {title} {\enquote {\bibinfo {title} {{Viscosity in strongly
  interacting quantum field theories from black hole physics}},}\ }\href
  {\doibase 10.1103/PhysRevLett.94.111601} {\bibfield  {journal} {\bibinfo
  {journal} {Phys.Rev.Lett.}\ }\textbf {\bibinfo {volume} {94}},\ \bibinfo
  {pages} {111601} (\bibinfo {year} {2005})},\ \Eprint
  {http://arxiv.org/abs/hep-th/0405231} {arXiv:hep-th/0405231 [hep-th]}
  \BibitemShut {NoStop}%
\bibitem [{\citenamefont {Romans}(1992)}]{Romans92}%
  \BibitemOpen
  \bibfield  {author} {\bibinfo {author} {\bibfnamefont {L.}~\bibnamefont
  {Romans}},\ }\bibfield  {title} {\enquote {\bibinfo {title} {{Supersymmetric,
  cold and lukewarm black holes in cosmological Einstein-Maxwell theory}},}\
  }\href {\doibase http://dx.doi.org/10.1016/0550-3213(92)90684-4} {\bibfield
  {journal} {\bibinfo  {journal} {Nuclear Physics B}\ }\textbf {\bibinfo
  {volume} {383}},\ \bibinfo {pages} {395 } (\bibinfo {year} {1992})},\ \Eprint
  {http://arxiv.org/abs/hep-th/9203018} {arXiv:hep-th/9203018 [hep-th]}
  \BibitemShut {NoStop}%
\bibitem [{\citenamefont {Chamblin}\ \emph {et~al.}(1999)\citenamefont
  {Chamblin}, \citenamefont {Emparan}, \citenamefont {Johnson},\ and\
  \citenamefont {Myers}}]{Myers99}%
  \BibitemOpen
  \bibfield  {author} {\bibinfo {author} {\bibfnamefont {A.}~\bibnamefont
  {Chamblin}}, \bibinfo {author} {\bibfnamefont {R.}~\bibnamefont {Emparan}},
  \bibinfo {author} {\bibfnamefont {C.~V.}\ \bibnamefont {Johnson}}, \ and\
  \bibinfo {author} {\bibfnamefont {R.~C.}\ \bibnamefont {Myers}},\ }\bibfield
  {title} {\enquote {\bibinfo {title} {{Charged AdS black holes and
  catastrophic holography}},}\ }\href {\doibase 10.1103/PhysRevD.60.064018}
  {\bibfield  {journal} {\bibinfo  {journal} {Phys. Rev. D}\ }\textbf {\bibinfo
  {volume} {60}},\ \bibinfo {pages} {064018} (\bibinfo {year} {1999})},\
  \Eprint {http://arxiv.org/abs/hep-th/9902170} {arXiv:hep-th/9902170 [hep-th]}
  \BibitemShut {NoStop}%
\bibitem [{\citenamefont {Faulkner}\ \emph
  {et~al.}(2011{\natexlab{a}})\citenamefont {Faulkner}, \citenamefont {Liu},
  \citenamefont {McGreevy},\ and\ \citenamefont {Vegh}}]{Faulkner09}%
  \BibitemOpen
  \bibfield  {author} {\bibinfo {author} {\bibfnamefont {T.}~\bibnamefont
  {Faulkner}}, \bibinfo {author} {\bibfnamefont {H.}~\bibnamefont {Liu}},
  \bibinfo {author} {\bibfnamefont {J.}~\bibnamefont {McGreevy}}, \ and\
  \bibinfo {author} {\bibfnamefont {D.}~\bibnamefont {Vegh}},\ }\bibfield
  {title} {\enquote {\bibinfo {title} {{Emergent quantum criticality, Fermi
  surfaces, and AdS(2)}},}\ }\href {\doibase 10.1103/PhysRevD.83.125002}
  {\bibfield  {journal} {\bibinfo  {journal} {Phys. Rev. D}\ }\textbf {\bibinfo
  {volume} {83}},\ \bibinfo {pages} {125002} (\bibinfo {year}
  {2011}{\natexlab{a}})},\ \Eprint {http://arxiv.org/abs/0907.2694}
  {arXiv:0907.2694 [hep-th]} \BibitemShut {NoStop}%
\bibitem [{\citenamefont {Cubrovic}\ \emph {et~al.}(2009)\citenamefont
  {Cubrovic}, \citenamefont {Zaanen},\ and\ \citenamefont {Schalm}}]{Zaanen}%
  \BibitemOpen
  \bibfield  {author} {\bibinfo {author} {\bibfnamefont {M.}~\bibnamefont
  {Cubrovic}}, \bibinfo {author} {\bibfnamefont {J.}~\bibnamefont {Zaanen}}, \
  and\ \bibinfo {author} {\bibfnamefont {K.}~\bibnamefont {Schalm}},\
  }\bibfield  {title} {\enquote {\bibinfo {title} {{String Theory, Quantum
  Phase Transitions and the Emergent Fermi-Liquid}},}\ }\href {\doibase
  10.1126/science.1174962} {\bibfield  {journal} {\bibinfo  {journal}
  {Science}\ }\textbf {\bibinfo {volume} {325}},\ \bibinfo {pages} {439}
  (\bibinfo {year} {2009})},\ \Eprint {http://arxiv.org/abs/0904.1993}
  {arXiv:0904.1993 [hep-th]} \BibitemShut {NoStop}%
\bibitem [{\citenamefont {Sen}(2005)}]{Sen05}%
  \BibitemOpen
  \bibfield  {author} {\bibinfo {author} {\bibfnamefont {A.}~\bibnamefont
  {Sen}},\ }\bibfield  {title} {\enquote {\bibinfo {title} {{Black hole entropy
  function and the attractor mechanism in higher derivative gravity}},}\ }\href
  {\doibase 10.1088/1126-6708/2005/09/038} {\bibfield  {journal} {\bibinfo
  {journal} {JHEP}\ }\textbf {\bibinfo {volume} {0509}},\ \bibinfo {pages}
  {038} (\bibinfo {year} {2005})},\ \Eprint
  {http://arxiv.org/abs/hep-th/0506177} {arXiv:hep-th/0506177 [hep-th]}
  \BibitemShut {NoStop}%
\bibitem [{\citenamefont {Sen}(2008)}]{Sen08}%
  \BibitemOpen
  \bibfield  {author} {\bibinfo {author} {\bibfnamefont {A.}~\bibnamefont
  {Sen}},\ }\bibfield  {title} {\enquote {\bibinfo {title} {{Entropy Function
  and AdS(2) / CFT(1) Correspondence}},}\ }\href {\doibase
  10.1088/1126-6708/2008/11/075} {\bibfield  {journal} {\bibinfo  {journal}
  {JHEP}\ }\textbf {\bibinfo {volume} {0811}},\ \bibinfo {pages} {075}
  (\bibinfo {year} {2008})},\ \Eprint {http://arxiv.org/abs/0805.0095}
  {arXiv:0805.0095 [hep-th]} \BibitemShut {NoStop}%
\bibitem [{\citenamefont {Wald}(1993)}]{Wald93}%
  \BibitemOpen
  \bibfield  {author} {\bibinfo {author} {\bibfnamefont {R.~M.}\ \bibnamefont
  {Wald}},\ }\bibfield  {title} {\enquote {\bibinfo {title} {{Black hole
  entropy is the Noether charge}},}\ }\href {\doibase
  10.1103/PhysRevD.48.R3427} {\bibfield  {journal} {\bibinfo  {journal} {\prd}\
  }\textbf {\bibinfo {volume} {48}},\ \bibinfo {pages} {3427} (\bibinfo {year}
  {1993})},\ \Eprint {http://arxiv.org/abs/gr-qc/9307038} {arXiv:gr-qc/9307038
  [gr-qc]} \BibitemShut {NoStop}%
\bibitem [{\citenamefont {Jacobson}\ \emph
  {et~al.}(1994{\natexlab{a}})\citenamefont {Jacobson}, \citenamefont {Kang},\
  and\ \citenamefont {Myers}}]{Myers93}%
  \BibitemOpen
  \bibfield  {author} {\bibinfo {author} {\bibfnamefont {T.}~\bibnamefont
  {Jacobson}}, \bibinfo {author} {\bibfnamefont {G.}~\bibnamefont {Kang}}, \
  and\ \bibinfo {author} {\bibfnamefont {R.~C.}\ \bibnamefont {Myers}},\
  }\bibfield  {title} {\enquote {\bibinfo {title} {{On black hole entropy}},}\
  }\href {\doibase 10.1103/PhysRevD.49.6587} {\bibfield  {journal} {\bibinfo
  {journal} {\prd}\ }\textbf {\bibinfo {volume} {49}},\ \bibinfo {pages} {6587}
  (\bibinfo {year} {1994}{\natexlab{a}})},\ \Eprint
  {http://arxiv.org/abs/gr-qc/9312023} {arXiv:gr-qc/9312023 [gr-qc]}
  \BibitemShut {NoStop}%
\bibitem [{\citenamefont {Iyer}\ and\ \citenamefont {Wald}(1994)}]{Wald94}%
  \BibitemOpen
  \bibfield  {author} {\bibinfo {author} {\bibfnamefont {V.}~\bibnamefont
  {Iyer}}\ and\ \bibinfo {author} {\bibfnamefont {R.~M.}\ \bibnamefont
  {Wald}},\ }\bibfield  {title} {\enquote {\bibinfo {title} {{Some properties
  of Noether charge and a proposal for dynamical black hole entropy}},}\ }\href
  {\doibase 10.1103/PhysRevD.50.846} {\bibfield  {journal} {\bibinfo  {journal}
  {\prd}\ }\textbf {\bibinfo {volume} {50}},\ \bibinfo {pages} {846} (\bibinfo
  {year} {1994})},\ \Eprint {http://arxiv.org/abs/gr-qc/9403028}
  {arXiv:gr-qc/9403028 [gr-qc]} \BibitemShut {NoStop}%
\bibitem [{\citenamefont {Jacobson}\ \emph
  {et~al.}(1994{\natexlab{b}})\citenamefont {Jacobson}, \citenamefont {Kang},\
  and\ \citenamefont {Myers}}]{Myers94}%
  \BibitemOpen
  \bibfield  {author} {\bibinfo {author} {\bibfnamefont {T.}~\bibnamefont
  {Jacobson}}, \bibinfo {author} {\bibfnamefont {G.}~\bibnamefont {Kang}}, \
  and\ \bibinfo {author} {\bibfnamefont {R.~C.}\ \bibnamefont {Myers}},\
  }\bibfield  {title} {\enquote {\bibinfo {title} {{Black hole entropy in
  higher curvature gravity}},}\ }\href@noop {} {\  (\bibinfo {year}
  {1994}{\natexlab{b}})},\ \Eprint {http://arxiv.org/abs/gr-qc/9502009}
  {arXiv:gr-qc/9502009 [gr-qc]} \BibitemShut {NoStop}%
\bibitem [{\citenamefont {Johnstone}\ \emph {et~al.}(2013)\citenamefont
  {Johnstone}, \citenamefont {Sheikh-Jabbari}, \citenamefont {Simon},\ and\
  \citenamefont {Yavartanoo}}]{Shahin13}%
  \BibitemOpen
  \bibfield  {author} {\bibinfo {author} {\bibfnamefont {M.}~\bibnamefont
  {Johnstone}}, \bibinfo {author} {\bibfnamefont {M.~M.}\ \bibnamefont
  {Sheikh-Jabbari}}, \bibinfo {author} {\bibfnamefont {J.}~\bibnamefont
  {Simon}}, \ and\ \bibinfo {author} {\bibfnamefont {H.}~\bibnamefont
  {Yavartanoo}},\ }\bibfield  {title} {\enquote {\bibinfo {title} {{Extremal
  black holes and the first law of thermodynamics}},}\ }\href {\doibase
  10.1103/PhysRevD.88.101503} {\bibfield  {journal} {\bibinfo  {journal}
  {\prd}\ }\textbf {\bibinfo {volume} {88}},\ \bibinfo {pages} {101503}
  (\bibinfo {year} {2013})},\ \Eprint {http://arxiv.org/abs/1305.3157}
  {arXiv:1305.3157 [hep-th]} \BibitemShut {NoStop}%
\bibitem [{\citenamefont {Sachdev}(2010{\natexlab{a}})}]{SS10}%
  \BibitemOpen
  \bibfield  {author} {\bibinfo {author} {\bibfnamefont {S.}~\bibnamefont
  {Sachdev}},\ }\bibfield  {title} {\enquote {\bibinfo {title} {{Holographic
  metals and the fractionalized Fermi liquid}},}\ }\href {\doibase
  10.1103/PhysRevLett.105.151602} {\bibfield  {journal} {\bibinfo  {journal}
  {Phys. Rev. Lett.}\ }\textbf {\bibinfo {volume} {105}},\ \bibinfo {pages}
  {151602} (\bibinfo {year} {2010}{\natexlab{a}})},\ \Eprint
  {http://arxiv.org/abs/1006.3794} {arXiv:1006.3794 [hep-th]} \BibitemShut
  {NoStop}%
\bibitem [{\citenamefont {Sachdev}(2010{\natexlab{b}})}]{SS10b}%
  \BibitemOpen
  \bibfield  {author} {\bibinfo {author} {\bibfnamefont {S.}~\bibnamefont
  {Sachdev}},\ }\bibfield  {title} {\enquote {\bibinfo {title} {{Strange metals
  and the AdS/CFT correspondence}},}\ }\href {\doibase
  10.1088/1742-5468/2010/11/P11022} {\bibfield  {journal} {\bibinfo  {journal}
  {J. Stat. Mech.}\ }\textbf {\bibinfo {volume} {1011}},\ \bibinfo {pages}
  {P11022} (\bibinfo {year} {2010}{\natexlab{b}})},\ \Eprint
  {http://arxiv.org/abs/1010.0682} {arXiv:1010.0682 [cond-mat.str-el]}
  \BibitemShut {NoStop}%
\bibitem [{\citenamefont {{Kitaev}}(2015)}]{AK15}%
  \BibitemOpen
  \bibfield  {author} {\bibinfo {author} {\bibfnamefont {A.~Y.}\ \bibnamefont
  {{Kitaev}}},\ }\bibfield  {title} {\enquote {\bibinfo {title} {{Talks at
  KITP, University of California, Santa Barbara}},}\ }\href
  {http://online.kitp.ucsb.edu/online/entangled15/} {\bibfield  {journal}
  {\bibinfo  {journal} {Entanglement in Strongly-Correlated Quantum Matter}\ }
  (\bibinfo {year} {2015})}\BibitemShut {NoStop}%
\bibitem [{\citenamefont {Iqbal}\ and\ \citenamefont {Liu}(2009)}]{Iqbal09}%
  \BibitemOpen
  \bibfield  {author} {\bibinfo {author} {\bibfnamefont {N.}~\bibnamefont
  {Iqbal}}\ and\ \bibinfo {author} {\bibfnamefont {H.}~\bibnamefont {Liu}},\
  }\bibfield  {title} {\enquote {\bibinfo {title} {{Real-time response in
  AdS/CFT with application to spinors}},}\ }\href {\doibase
  10.1002/prop.200900057} {\bibfield  {journal} {\bibinfo  {journal} {Fortsch.
  Phys.}\ }\textbf {\bibinfo {volume} {57}},\ \bibinfo {pages} {367} (\bibinfo
  {year} {2009})},\ \Eprint {http://arxiv.org/abs/0903.2596} {arXiv:0903.2596
  [hep-th]} \BibitemShut {NoStop}%
\bibitem [{\citenamefont {Faulkner}\ \emph
  {et~al.}(2011{\natexlab{b}})\citenamefont {Faulkner}, \citenamefont {Iqbal},
  \citenamefont {Liu}, \citenamefont {McGreevy},\ and\ \citenamefont
  {Vegh}}]{Faulkner11}%
  \BibitemOpen
  \bibfield  {author} {\bibinfo {author} {\bibfnamefont {T.}~\bibnamefont
  {Faulkner}}, \bibinfo {author} {\bibfnamefont {N.}~\bibnamefont {Iqbal}},
  \bibinfo {author} {\bibfnamefont {H.}~\bibnamefont {Liu}}, \bibinfo {author}
  {\bibfnamefont {J.}~\bibnamefont {McGreevy}}, \ and\ \bibinfo {author}
  {\bibfnamefont {D.}~\bibnamefont {Vegh}},\ }\bibfield  {title} {\enquote
  {\bibinfo {title} {{Holographic non-Fermi liquid fixed points}},}\ }\href
  {\doibase 10.1098/rsta.2010.0354} {\bibfield  {journal} {\bibinfo  {journal}
  {Phil. Trans. Roy. Soc. A}\ }\textbf {\bibinfo {volume} {369}},\ \bibinfo
  {pages} {1640} (\bibinfo {year} {2011}{\natexlab{b}})},\ \Eprint
  {http://arxiv.org/abs/1101.0597} {arXiv:1101.0597 [hep-th]} \BibitemShut
  {NoStop}%
\bibitem [{\citenamefont {Gibbons}\ and\ \citenamefont
  {Hawking}(1977)}]{Gibbons77}%
  \BibitemOpen
  \bibfield  {author} {\bibinfo {author} {\bibfnamefont {G.~W.}\ \bibnamefont
  {Gibbons}}\ and\ \bibinfo {author} {\bibfnamefont {S.~W.}\ \bibnamefont
  {Hawking}},\ }\bibfield  {title} {\enquote {\bibinfo {title} {Action
  integrals and partition functions in quantum gravity},}\ }\href {\doibase
  10.1103/PhysRevD.15.2752} {\bibfield  {journal} {\bibinfo  {journal} {Phys.
  Rev. D}\ }\textbf {\bibinfo {volume} {15}},\ \bibinfo {pages} {2752}
  (\bibinfo {year} {1977})}\BibitemShut {NoStop}%
\bibitem [{\citenamefont {Hartman}\ and\ \citenamefont
  {Strominger}(2009)}]{Hartman08}%
  \BibitemOpen
  \bibfield  {author} {\bibinfo {author} {\bibfnamefont {T.}~\bibnamefont
  {Hartman}}\ and\ \bibinfo {author} {\bibfnamefont {A.}~\bibnamefont
  {Strominger}},\ }\bibfield  {title} {\enquote {\bibinfo {title} {{Central
  Charge for AdS(2) Quantum Gravity}},}\ }\href {\doibase
  10.1088/1126-6708/2009/04/026} {\bibfield  {journal} {\bibinfo  {journal}
  {JHEP}\ }\textbf {\bibinfo {volume} {0904}},\ \bibinfo {pages} {026}
  (\bibinfo {year} {2009})},\ \Eprint {http://arxiv.org/abs/0803.3621}
  {arXiv:0803.3621 [hep-th]} \BibitemShut {NoStop}%
\bibitem [{\citenamefont {Bardeen}\ \emph {et~al.}(1973)\citenamefont
  {Bardeen}, \citenamefont {Carter},\ and\ \citenamefont
  {Hawking}}]{Bardeen73}%
  \BibitemOpen
  \bibfield  {author} {\bibinfo {author} {\bibfnamefont {J.~M.}\ \bibnamefont
  {Bardeen}}, \bibinfo {author} {\bibfnamefont {B.}~\bibnamefont {Carter}}, \
  and\ \bibinfo {author} {\bibfnamefont {S.~W.}\ \bibnamefont {Hawking}},\
  }\bibfield  {title} {\enquote {\bibinfo {title} {The four laws of black hole
  mechanics},}\ }\href {\doibase 10.1007/BF01645742} {\bibfield  {journal}
  {\bibinfo  {journal} {Communications in Mathematical Physics}\ }\textbf
  {\bibinfo {volume} {31}},\ \bibinfo {pages} {161} (\bibinfo {year}
  {1973})}\BibitemShut {NoStop}%
\bibitem [{\citenamefont {Strominger}\ and\ \citenamefont {Vafa}(1996)}]{ASCV}%
  \BibitemOpen
  \bibfield  {author} {\bibinfo {author} {\bibfnamefont {A.}~\bibnamefont
  {Strominger}}\ and\ \bibinfo {author} {\bibfnamefont {C.}~\bibnamefont
  {Vafa}},\ }\bibfield  {title} {\enquote {\bibinfo {title} {{Microscopic
  origin of the Bekenstein-Hawking entropy}},}\ }\href {\doibase
  10.1016/0370-2693(96)00345-0} {\bibfield  {journal} {\bibinfo  {journal}
  {Phys.Lett.}\ }\textbf {\bibinfo {volume} {B379}},\ \bibinfo {pages} {99}
  (\bibinfo {year} {1996})},\ \Eprint {http://arxiv.org/abs/hep-th/9601029}
  {arXiv:hep-th/9601029 [hep-th]} \BibitemShut {NoStop}%
\bibitem [{\citenamefont {Bekenstein}(1973)}]{JDB73}%
  \BibitemOpen
  \bibfield  {author} {\bibinfo {author} {\bibfnamefont {J.~D.}\ \bibnamefont
  {Bekenstein}},\ }\bibfield  {title} {\enquote {\bibinfo {title} {Black holes
  and entropy},}\ }\href {\doibase 10.1103/PhysRevD.7.2333} {\bibfield
  {journal} {\bibinfo  {journal} {Phys. Rev. D}\ }\textbf {\bibinfo {volume}
  {7}},\ \bibinfo {pages} {2333} (\bibinfo {year} {1973})}\BibitemShut
  {NoStop}%
\bibitem [{\citenamefont {Hawking}(1975)}]{SWH75}%
  \BibitemOpen
  \bibfield  {author} {\bibinfo {author} {\bibfnamefont {S.~W.}\ \bibnamefont
  {Hawking}},\ }\bibfield  {title} {\enquote {\bibinfo {title} {{Particle
  creation by black holes}},}\ }\href {\doibase 10.1007/BF02345020} {\bibfield
  {journal} {\bibinfo  {journal} {Communications in Mathematical Physics}\
  }\textbf {\bibinfo {volume} {43}},\ \bibinfo {pages} {199} (\bibinfo {year}
  {1975})}\BibitemShut {NoStop}%
\bibitem [{\citenamefont {{Georges}}\ \emph {et~al.}(2000)\citenamefont
  {{Georges}}, \citenamefont {{Parcollet}},\ and\ \citenamefont
  {{Sachdev}}}]{GPS99}%
  \BibitemOpen
  \bibfield  {author} {\bibinfo {author} {\bibfnamefont {A.}~\bibnamefont
  {{Georges}}}, \bibinfo {author} {\bibfnamefont {O.}~\bibnamefont
  {{Parcollet}}}, \ and\ \bibinfo {author} {\bibfnamefont {S.}~\bibnamefont
  {{Sachdev}}},\ }\bibfield  {title} {\enquote {\bibinfo {title} {{Mean Field
  Theory of a Quantum Heisenberg Spin Glass}},}\ }\href {\doibase
  10.1103/PhysRevLett.85.840} {\bibfield  {journal} {\bibinfo  {journal}
  {\prl}\ }\textbf {\bibinfo {volume} {85}},\ \bibinfo {pages} {840} (\bibinfo
  {year} {2000})},\ \Eprint {http://arxiv.org/abs/cond-mat/9909239}
  {cond-mat/9909239} \BibitemShut {NoStop}%
\bibitem [{\citenamefont {{Arrachea}}\ and\ \citenamefont
  {{Rozenberg}}(2002)}]{MJR02}%
  \BibitemOpen
  \bibfield  {author} {\bibinfo {author} {\bibfnamefont {L.}~\bibnamefont
  {{Arrachea}}}\ and\ \bibinfo {author} {\bibfnamefont {M.~J.}\ \bibnamefont
  {{Rozenberg}}},\ }\bibfield  {title} {\enquote {\bibinfo {title}
  {{Infinite-range quantum random Heisenberg magnet}},}\ }\href {\doibase
  10.1103/PhysRevB.65.224430} {\bibfield  {journal} {\bibinfo  {journal}
  {\prb}\ }\textbf {\bibinfo {volume} {65}},\ \bibinfo {eid} {224430} (\bibinfo
  {year} {2002})},\ \Eprint {http://arxiv.org/abs/cond-mat/0203537}
  {cond-mat/0203537} \BibitemShut {NoStop}%
\bibitem [{\citenamefont {{Camjayi}}\ and\ \citenamefont
  {{Rozenberg}}(2003)}]{MJR03}%
  \BibitemOpen
  \bibfield  {author} {\bibinfo {author} {\bibfnamefont {A.}~\bibnamefont
  {{Camjayi}}}\ and\ \bibinfo {author} {\bibfnamefont {M.~J.}\ \bibnamefont
  {{Rozenberg}}},\ }\bibfield  {title} {\enquote {\bibinfo {title} {{Quantum
  and Thermal Fluctuations in the SU(N) Heisenberg Spin-Glass Model near the
  Quantum Critical Point}},}\ }\href {\doibase 10.1103/PhysRevLett.90.217202}
  {\bibfield  {journal} {\bibinfo  {journal} {\prl}\ }\textbf {\bibinfo
  {volume} {90}},\ \bibinfo {eid} {217202} (\bibinfo {year} {2003})},\ \Eprint
  {http://arxiv.org/abs/cond-mat/0210407} {cond-mat/0210407} \BibitemShut
  {NoStop}%
\bibitem [{\citenamefont {Faulkner}\ and\ \citenamefont
  {Polchinski}(2011)}]{TFJP11}%
  \BibitemOpen
  \bibfield  {author} {\bibinfo {author} {\bibfnamefont {T.}~\bibnamefont
  {Faulkner}}\ and\ \bibinfo {author} {\bibfnamefont {J.}~\bibnamefont
  {Polchinski}},\ }\bibfield  {title} {\enquote {\bibinfo {title}
  {{Semi-Holographic Fermi Liquids}},}\ }\href {\doibase
  10.1007/JHEP06(2011)012} {\bibfield  {journal} {\bibinfo  {journal} {JHEP}\
  }\textbf {\bibinfo {volume} {1106}},\ \bibinfo {pages} {012} (\bibinfo {year}
  {2011})},\ \Eprint {http://arxiv.org/abs/1001.5049} {arXiv:1001.5049
  [hep-th]} \BibitemShut {NoStop}%
\bibitem [{\citenamefont {{Metlitski}}\ and\ \citenamefont
  {{Sachdev}}(2010)}]{MS10}%
  \BibitemOpen
  \bibfield  {author} {\bibinfo {author} {\bibfnamefont {M.~A.}\ \bibnamefont
  {{Metlitski}}}\ and\ \bibinfo {author} {\bibfnamefont {S.}~\bibnamefont
  {{Sachdev}}},\ }\bibfield  {title} {\enquote {\bibinfo {title} {{Quantum
  phase transitions of metals in two spatial dimensions. I.~Ising-nematic
  order}},}\ }\href {\doibase 10.1103/PhysRevB.82.075127} {\bibfield  {journal}
  {\bibinfo  {journal} {\prb}\ }\textbf {\bibinfo {volume} {82}},\ \bibinfo
  {eid} {075127} (\bibinfo {year} {2010})},\ \Eprint
  {http://arxiv.org/abs/1001.1153} {arXiv:1001.1153 [cond-mat.str-el]}
  \BibitemShut {NoStop}%
\bibitem [{\citenamefont {{Mross}}\ \emph {et~al.}(2010)\citenamefont
  {{Mross}}, \citenamefont {{McGreevy}}, \citenamefont {{Liu}},\ and\
  \citenamefont {{Senthil}}}]{MMLS10}%
  \BibitemOpen
  \bibfield  {author} {\bibinfo {author} {\bibfnamefont {D.~F.}\ \bibnamefont
  {{Mross}}}, \bibinfo {author} {\bibfnamefont {J.}~\bibnamefont {{McGreevy}}},
  \bibinfo {author} {\bibfnamefont {H.}~\bibnamefont {{Liu}}}, \ and\ \bibinfo
  {author} {\bibfnamefont {T.}~\bibnamefont {{Senthil}}},\ }\bibfield  {title}
  {\enquote {\bibinfo {title} {{Controlled expansion for certain
  non-Fermi-liquid metals}},}\ }\href {\doibase 10.1103/PhysRevB.82.045121}
  {\bibfield  {journal} {\bibinfo  {journal} {\prb}\ }\textbf {\bibinfo
  {volume} {82}},\ \bibinfo {eid} {045121} (\bibinfo {year} {2010})},\ \Eprint
  {http://arxiv.org/abs/1003.0894} {arXiv:1003.0894 [cond-mat.str-el]}
  \BibitemShut {NoStop}%
\bibitem [{\citenamefont {DeWolfe}\ \emph {et~al.}(2012)\citenamefont
  {DeWolfe}, \citenamefont {Gubser},\ and\ \citenamefont {Rosen}}]{Gubser11}%
  \BibitemOpen
  \bibfield  {author} {\bibinfo {author} {\bibfnamefont {O.}~\bibnamefont
  {DeWolfe}}, \bibinfo {author} {\bibfnamefont {S.~S.}\ \bibnamefont {Gubser}},
  \ and\ \bibinfo {author} {\bibfnamefont {C.}~\bibnamefont {Rosen}},\
  }\bibfield  {title} {\enquote {\bibinfo {title} {{Fermi Surfaces in Maximal
  Gauged Supergravity}},}\ }\href {\doibase 10.1103/PhysRevLett.108.251601}
  {\bibfield  {journal} {\bibinfo  {journal} {Phys.Rev.Lett.}\ }\textbf
  {\bibinfo {volume} {108}},\ \bibinfo {pages} {251601} (\bibinfo {year}
  {2012})},\ \Eprint {http://arxiv.org/abs/1112.3036} {arXiv:1112.3036
  [hep-th]} \BibitemShut {NoStop}%
\bibitem [{\citenamefont {Shenker}\ and\ \citenamefont
  {Stanford}(2014)}]{Shenker13}%
  \BibitemOpen
  \bibfield  {author} {\bibinfo {author} {\bibfnamefont {S.~H.}\ \bibnamefont
  {Shenker}}\ and\ \bibinfo {author} {\bibfnamefont {D.}~\bibnamefont
  {Stanford}},\ }\bibfield  {title} {\enquote {\bibinfo {title} {{Black holes
  and the butterfly effect}},}\ }\href {\doibase 10.1007/JHEP03(2014)067}
  {\bibfield  {journal} {\bibinfo  {journal} {JHEP}\ }\textbf {\bibinfo
  {volume} {1403}},\ \bibinfo {pages} {067} (\bibinfo {year} {2014})},\ \Eprint
  {http://arxiv.org/abs/1306.0622} {arXiv:1306.0622 [hep-th]} \BibitemShut
  {NoStop}%
\bibitem [{\citenamefont {Maldacena}\ \emph {et~al.}(2015)\citenamefont
  {Maldacena}, \citenamefont {Shenker},\ and\ \citenamefont
  {Stanford}}]{Shenker15}%
  \BibitemOpen
  \bibfield  {author} {\bibinfo {author} {\bibfnamefont {J.}~\bibnamefont
  {Maldacena}}, \bibinfo {author} {\bibfnamefont {S.~H.}\ \bibnamefont
  {Shenker}}, \ and\ \bibinfo {author} {\bibfnamefont {D.}~\bibnamefont
  {Stanford}},\ }\bibfield  {title} {\enquote {\bibinfo {title} {{A bound on
  chaos}},}\ }\href@noop {} {\  (\bibinfo {year} {2015})},\ \Eprint
  {http://arxiv.org/abs/1503.01409} {arXiv:1503.01409 [hep-th]} \BibitemShut
  {NoStop}%
\bibitem [{\citenamefont {Luttinger}\ and\ \citenamefont {Ward}(1960)}]{LW60}%
  \BibitemOpen
  \bibfield  {author} {\bibinfo {author} {\bibfnamefont {J.~M.}\ \bibnamefont
  {Luttinger}}\ and\ \bibinfo {author} {\bibfnamefont {J.~C.}\ \bibnamefont
  {Ward}},\ }\bibfield  {title} {\enquote {\bibinfo {title} {{Ground-State
  Energy of a Many-Fermion System. II}},}\ }\href {\doibase
  10.1103/PhysRev.118.1417} {\bibfield  {journal} {\bibinfo  {journal} {Phys.
  Rev.}\ }\textbf {\bibinfo {volume} {118}},\ \bibinfo {pages} {1417} (\bibinfo
  {year} {1960})}\BibitemShut {NoStop}%
\end{thebibliography}%
\end{document}